\documentclass[twocolumn,epjc3]{svjour3}          

\usepackage{graphicx}
\usepackage{dcolumn}
\usepackage{bm}
\usepackage{amsmath,multirow}
\usepackage{amssymb}
\usepackage{subfigure}
\usepackage{color}
\usepackage{mathrsfs}
\usepackage[section]{placeins}

\RequirePackage[T1]{fontenc}
\smartqed  

\journalname{EPJC}

\begin{document}

\title{Electromagnetic field and chaotic charged-particle motion around hairy black holes in Horndeski gravity}

\author{
Wenfu Cao$^{a,1,2}$\and Xin Wu$^{c,1,2}$ \and
Jun Lyu$^{d,3}$}

\thankstext{e1}{e-mail: M440121503$@$sues.edu.cn}
\thankstext{e2}{e-mail: wuxin\_1134$@$sina.com}
\thankstext{e3}{e-mail: junlyu2023$@$163.com}
\institute{School of Mathematics, Physics and Statistics, Shanghai
University of Engineering Science, Shanghai 201620, China
\label{addr1} \and Center of Application and Research of
Computational Physics, Shanghai University of Engineering Science,
Shanghai, China \label{addr2} \and School of Mathematics and
Statistics, Yunnan University, Kunming 650500, China \label{addr3}
}

\date{Received: date / Accepted: date}

\maketitle

\begin{abstract}

The Wald vector potential is an exact solution of the source-less
Maxwell equations regarding an electromagnetic field of a vacuum
uncharged black hole like the Kerr background black hole in an
asymptotically uniform magnetic field. However, it is not if the
black hole is a nonvacuum solution
in a theory of modified gravity with extra fields or
a charged Kerr-Newman spacetime. To satisfy the source-less Maxwell
equations in this case, the Wald vector potential must be modified
and generalized appropriately. Following this idea, we derive
an expression for the vector potential of an electromagnetic field surrounding
a hairy black hole in the Horndeski modified gravity theory.
Explicit symplectic integrators with excellent long-term
behaviour are used to simulate the motion of charged particles around the
hairy black hole immersed in the external magnetic field.  The
recurrence plot method based on the recurrence quantification
analysis uses diagonal structures parallel to the main diagonal to
show  regular dynamics, but adopts no diagonal structures to
indicate chaotic dynamics. The method is efficient to detect chaos from
order in the curved spacetime, as the Poincar\'{e} map and the fast Lyapunov indicator are.

\end{abstract}

\tableofcontents

\section{Introduction}

The existence of black holes is an important prediction of the
theory of  general relativity. It has been confirmed frequently
through the detection of gravitational waves by LIGO [1] and the
observations of the shadow images of supermassive black holes M87*
and Sgr A* by the Event Horizon Telescope (EHT) [2]. However,
the theory of general relativity does not admit  the
emergence of dark energy responsible for  the apparent
accelerating expansion of the Universe. To cure the limits of
general relativity, modified or extended gravity theories are
necessarily given. Some of them are scalar-tensor theories [3-5],
Einstein-\ae ther theories [6-8],
quantum gravity theories [9,10] and  Einstein-scalar-Gauss-Bonnet
theories [11].

Several observations support  the presence of magnetic fields
around astrophysical black holes [12-14]. A possible generation of
magnetic fields is due to the dynamo mechanism in the plasma of
accretion disks around the central black holes [12]. The magnetic
fields in the vicinity of the black holes  is generally believed
to transfer the energy from the accretion disc to relativistic
jets. In this sense, they are helpful for one to understand the
formation and energetics of the black hole jets. However, the
supermassive black hole at the centre of the Galaxy is surrounded
by a strong magnetic field independent of an accretion disc [14].
In spite of the different claim, the existence of an
asymptotically uniform test magnetic field in the vicinity of a
black hole was shown at large enough distance to the magnetar
[15]. The obtainment of such an external, large-scale
electromagnetic field is based on a Wald solution [16-19]. The
Wald solution requires that the  black hole should be uncharged
like the Kerr background black hole or the Schwarzschild one. In addition to this requirement,
the external magnetic field should be such a weak field that its  strength
is much smaller than $10^{19}M_{\odot}/M$ Gauss [20], where
$M_{\odot}$ and $M$ respectively correspond to the masses of the
Sun and black hole. When the black hole is electrically charged,
nonvacuum like  the Kerr-Newman black hole or the
Reissner-Nordstr\"{o}m (RN) one, the Wald vector potential does
not exactly satisfy the source-less Maxwell equations, as was
claimed by Azreg-A\"{i}nou [21]. This result is also suitable for
the case of a uniform magnetic field near a nonvacuum black hole
of modified gravity with  extra fields. To satisfy the source-less
Maxwell equations in the two cases, the Wald vector potential has
to be modified appropriately. Although relatively weak external
magnetic fields have a negligible effect on the spacetime
background, they can strongly influence the motion of charged
particles in the vicinity of the black hole horizon. The charged
particle chaotic motions induced by the  magnetic fields have
appeared in a number of studies [22-33]. If the external magnetic
fields have their strengths close to $10^{19}M_{\odot}/M$ Gauss,
they not only change the metric tensor of the black hole spacetime
[34-37] but also influence the motion of neutral or charged
particles in the vicinity of the black hole horizon. The chaotic
motion of neutral particles can be found in several references
[38-43].

Although a Hamiltonian system for the description of neutral or charged particle motion
in the vicinity of a black hole immersed in an external electromagnetic field is inseparable
and exhibits chaotic character in most cases,
it or its time-transformed version may have more than two explicitly integrable splitting pieces. This brings a chance for the construction of
explicit symplectic integrators in curved spacetimes [44-49]. The symplectic integrators preserve the symplectic structures of Hamiltonian dynamics,
and show no secular drifts in errors of the integrals of motion [50,51]. Thus, they are particularly adapted to mimicking
the long-term dynamical evolution of Hamiltonian systems.
The explicit symplectic methods have an advantage over the implicit ones [52,53] in computational efficiency.

In addition to reliable numerical integration methods such as the
symplectic integrators, efficient chaos detection methods are
necessary to discriminate between order and chaos of Hamiltonian
dynamics. The Poincar\'{e} map is a common  chaos detection method
applied to a conservative system with two degrees of freedom. The
maximal Lyapunov exponent (mLE) is also a common  chaos detection
method used in a system with any dimensions. The fast Lyapunov
indicator (FLI) [54] as a variant of the mLE is a quicker indicator
to detect the chaotical behaviour. The mLE and FLI were developed
as those independent of the choice of spacetime coordinates in
general relativity [55,56]. There are other chaos detection methods,
which include the smaller alignment index (SALI) [57] and its
generalized alignment index (GALI) [58], the 0-1 binary test
correlation method [59], the recurrence plot (RP) method based on
the recurrence quantification analysis [23,60-63] and so on. Although
these  techniques have been shown to be very efficient to detect
the chaos onset in many Newtonian gravitational systems, they have
few applications in the context of relativistic gravitational
problems. The RP method was applied to identify the transition
between different dynamical regimes in the Kerr black hole
background immersed in a weak, asymptotically uniform magnetic
field [23,62]. The SALI was used to trace the chaotic motion of charged
particles around a deformed Schwarzschild black hole with an
external magnetic field [64]. Recently, the 0-1 binary test
correlation method was employed to identify chaos in magnetized
Kerr-Newman spacetimes [43].

In the present paper, we use the RP method combined
with an explicit symplectic method to study the chaotic dynamics of charged
test particles around a hairy black hole with an external magnetic
field in the Horndeski gravity [65]. Because the Horndeski gravity
is a very general scalar-tensor theory belonging to a theory of
modified gravity [66-68], we  give the vector potential of electromagnetic
field by modifying the Wald vector potential in Section 2. Then we
explore the dynamical properties of charged test particles in
Section 3. Finally, our main results are concluded in Section 4.

\section{Electromagnetic field
of hairy black hole in Horndeski gravity}\label{section2}

There have been many interesting solutions describing hairy black
holes in literature (see e.g. [65-72]). Here, we focus on a hairy
black hole spacetime in Horndeski gravity [65-68]. Above all, the
expressions for the vector potential and electromagnetic field
around the hairy black hole are given.

\subsection{Hairy black hole solution in sqrt quartic Horndeski gravity}

Horndeski gravity [65] is a very general scalar-tensor theory, which belongs to a theory of modified gravity.
A scalar field $\phi$ in the theory has a shift
symmetry $\phi \to \phi + \text{const}$. We focus on one black hole solution in the class of Horndeski theory,
which is obtained from the action [66-68]
\begin{equation}
S=\int \mathrm{d}^4 x \sqrt{-g}\left(\mathcal{L}_2+\mathcal{L}_4\right),
\end{equation}
where $\mathrm{d}^4 x$ is products  of infinitesimal spacetime coordinate elements (such as $dt$, $dx$, $dy$ and $dz$, or
$dt$, $dr$, $d\theta$ and $d\varphi$),
and $g$ is  the determinant of a metric matrix $g_{\alpha\beta}$.
In additon, $\mathcal{L}_2$ and $\mathcal{L}_4$ are two Lagrangian terms:
\begin{eqnarray}
\label{eq:2}
{\cal L}_2 &=& G_2(X),\\
\label{eq:4}
{\cal L}_4 &=& G_4(X) R + G_{4,X} \left [ (\Box \phi)^2 -(\nabla_\mu\nabla_\nu\phi)^2 \right ].
\end{eqnarray}
Here, $G_2$ and $G_4$ are two arbitrary functions of a canonical kinetic term $X =-\partial^\mu \phi \partial_\mu \phi / 2$.
$G_{4,X}$ corresponds to a derivative of $G_4$ with respect to $X$, and  $R$ stands for the Ricci scalar.
$(\Box \phi)^2=\left(\partial_{\mu} \partial^{\mu}\phi \right)^2$,
and $\left(\nabla_\mu \nabla_\nu \phi\right)^2 = \nabla_\mu \nabla_\nu \phi \nabla^\nu \nabla^\mu \phi$.
For example, $G_2$ and $G_4$ are chosen as
\begin{equation}
G_2=\eta X, \quad G_4=1 / (16 \pi)+\beta \sqrt{-X},
\end{equation}
where $\eta$ and $\beta$ represent two dimensionless parameters.
It is clear that $G_2$ depends on the canonical kinetic term only.
$G_4$ has the constant term $1/(16 \pi)$ associated with an Einstein-Hilbert piece in the action
and  the $X$-dependence term yielding
a contribution to the Noether current that is independent of the scalar field $\phi$.
Because of the expression of the Lagrangian term $\mathcal{L}_4$, the theory given by the action (1)
was referred to as a "sqrt quartic Horndeski gravity" in [66,67].

Using the action (1), the authors of [66,67] wrote a hairy black hole solution in Boyer-Lindquist
coordinates as
\begin{eqnarray}
ds^{2} &=& g_{\mu\nu}dx^{\mu}dx^{\nu} \nonumber \\
&=& -(1-\frac{2M}{r}-\frac{8\pi \rho_{\mathrm{eff}}}{r^2})dt^{2} \nonumber
\\&& +(1-\frac{2M}{r}-\frac{8\pi  \rho_{\mathrm{eff}}}{r^2})^{-1} dr^{2}\nonumber
\\
& &+r^{2}d \theta^{2} +r^{2}\sin^{2} \theta d \varphi^2,
\label{eq:10}
\end{eqnarray}
where $\rho_{\mathrm{eff}} = \beta^2 / \eta$ is a scalar hairy parameter.
If the hairy parameter satisfies the condition
\begin{equation}
\rho_{\mathrm{eff}}\geq -M^2/(8\pi),
\end{equation}
the black hole solution has one or two event
horizons $r_{\pm}=M\pm\sqrt{M^2+8\pi\rho_{eff}}$. For
$\rho_{\mathrm{eff}}=- M^2/(8\pi)$, the metric (5) corresponds to
the black hole with one horizon  $r=M$. When
$-M^2/(8\pi)<\rho_{\mathrm{eff}}<0$, the metric  is the RN-like
black hole with two event horizons, where
$Q_{sc}=\sqrt{-8\pi\rho_{\mathrm{eff}}}$ is a scalar charge [73].
Although it is not an electric charge of the black hole, it has a
physical behaviour like  the electric charge effect of the black
hole. If the hairy parameter $\rho_{\mathrm{eff}}$ is nonzero, it
yields an extra force for the neutral black hole. When
$\rho_{\mathrm{eff}}<0$, the extra force leads to the acceleration
of positively charged particles in the increasing $\varphi$
direction, but to the deceleration of negatively charged
particles. In a word, the effect should be considered in the
vicinity of the event horizon. For $\rho_{\mathrm{eff}}=0$, the
metric becomes the Schwarzschild spacetime  with one horizon
$r=2M$. If $\rho_{\mathrm{eff}}>0$, the metric is unlike the RN
black hole but has two event horizons. However, the metric is no
longer any black hole solution when $\rho_{\mathrm{eff}}<
-M^2/(8\pi)$ because no event horizon can exist in this case.
Throughout this paper, the constant of gravity $G$ and the speed
of light $c$ take one geometric unit, $G=c=1$.

The hairy black hole spacetime (5) is spherically symmetric and asymptotically flat.
There are two Killing vectors $\xi^{\mu}_{t}=(1,0,0,0)$ and $\xi^{\mu}_{\varphi}=(0,0,0,1)$.
They determine the conserved energy $\bar{E}$ and orbital angular momentum $\bar{L}$ per unit mass of a test particle:
\begin{eqnarray}
\bar{E}&=&-u^{\mu}\xi_{t\mu}=-u^{\mu}g_{\mu\nu}\xi^{\nu}_{t} =-g_{tt}\dot{t}=-\bar{P}_t, \label{eq:11}\\
\bar{L} &=&u^{\mu}\xi_{\varphi\mu}=u^{\mu}g_{\mu\nu}\xi^{\nu}_{\varphi} = g_{\varphi\varphi}\dot{\varphi}=\bar{P}_{\varphi}, \label{eq:12}
\end{eqnarray}
where the dots denote the derivatives of the coordinates $t$ and $\varphi$ with respect to the proper time $\tau$,
i.e. two components of the 4-velocity $u^{\mu}=\dot{x}^{\mu}$.
$\bar{P}_t$ and $\bar{P}_{\varphi}$ are two components of the generalized 4-momentum $\bar{P}_{\mu}=g_{\mu\nu}\dot{x}^{\nu}$.
For the test particle with mass $m$ around the black hole, its motion can be described by a Hamiltonian system
\begin{eqnarray}
H &=& \frac{1}{2m} g^{\mu\nu}P_{\mu}P_{\nu} \nonumber \\
&=& -\frac{\text{E}^2}{2m \left(1-\frac{2M}{r}-\frac{8\pi  \rho _{\text{eff}}}{r^2}\right)}+\frac{L^2 \csc^2\theta}{2m r^2}  \nonumber \\
&& +\frac{P^2_r}{2m} \left(1-\frac{2M}{r}-\frac{8\pi  \rho _{\text{eff}}}{r^2}\right)+\frac{P_{\theta }^2}{2m r^2},
\end{eqnarray}
where $P_{\mu}=m\bar{P}_{\mu}$, $E=m\bar{E}$ and $L=m\bar{L}$. This Hamiltonian is a conserved quantity
\begin{equation}
H=-\frac{m}{2},
\end{equation}
which is due to the rest mass relation $g_{\mu\nu}\dot{x}^{\mu}\dot{x}^{\nu}=-1$.
There is a fourth constant of motion, which is obtained from the separation of variables in the Hamilton-Jacobi equation of the Hamiltonian system
(9). The fourth constant reads
\begin{eqnarray}
C_k &=& \frac{r^2\text{E}^2}{m \left(1-\frac{2M}{r}-\frac{8\pi  \rho _{\text{eff}}}{r^2}\right)}-mr^2  \nonumber \\
&& -\frac{r^2P^2_r}{m} \left(1-\frac{2M}{r}-\frac{8\pi  \rho _{\text{eff}}}{r^2}\right) \nonumber \\
&=&\frac{P_{\theta }^2}{m}+\frac{L^2 \csc^2\theta}{m},
\end{eqnarray}
where $C_k$ is the Carter-like constant.
Thus, the Hamiltonian system (9) is integrable and has formally analytical
solutions.

For simplicity, we give dimensionless operations to the Hamiltonian system (9) via scale transformations:
$t\rightarrow Mt$, $\tau\rightarrow M\tau$, $r\rightarrow Mr$, $\rho _{\text{eff}}\rightarrow M^2 \rho _{\text{eff}}$,
$E\rightarrow mE$, $L\rightarrow mML$,  $P_r\rightarrow mP_r$, $P_{\theta }\rightarrow mMP_{\theta }$,
$H\rightarrow mH$ and $C_k\rightarrow mM^2C_k$. In this way, the mass factors $m$ and $M$ are eliminated in all the above expressions.
That is, $P_{\mu}=\bar{P}_{\mu}$, $E=\bar{E}$, $L=\bar{L}$, $\cdots$. Hereafter, the scaled quantities
are still represented in terms of $P_{\mu}$, $E$, $L$, $\rho _{\text{eff}}$ and so on for convenience.
These operations give not only the simple expressions but also the links between the practical quantities
and the scaled ones. For instance, the transformation $\rho _{\text{eff,pra}}\rightarrow M^2 \rho _{\text{eff,sca}}$
means that the scaled value of the hairy parameter $\rho_{\mathrm{eff, sca}}$ corresponds to
its practical value $\rho _{\text{eff,pra}}$ equal to $M^2\rho_{\mathrm{eff, sca}}$.

\subsection{Electromagnetic field around the hairy black hole}

Wald [16] considered the solution for an electromagnetic field of a vacuum, stationary, axisymmetric
black hole immersed in a uniform magnetic field aligned with the axis of
symmetry of the black hole.
The solution is described by the vector potential
\begin{equation}
A^{\mu} =C_{t}(B)\xi^{\mu}_{t}+C_{\varphi}(B)\xi^{\mu}_{\varphi},
\end{equation}
where coefficients $C_{t}$ and $C_{\varphi}$ depend on the magnetic field strength $B$.
When the black hole is  neutral, the coefficients obtained from the Wald's solution are
\begin{equation}
C_{t}=aB, ~~~~ C_{\varphi}=\frac{B}{2},
\end{equation}
where $a$ is the rotation angular momentum of the black hole. If the black hole has
an electric charge $Q$, the coefficients are
\begin{equation}
C_{t}=aB+\frac{Q}{2M}, ~~~~ C_{\varphi}=\frac{B}{2}.
\end{equation}

Azreg-A\"{i}nou [21] pointed out that the potential (12) given by the coefficients (13) in the Kerr background metric
(including the Schwarzschild one) is an exact solution of the source-less Maxwell equations
\begin{eqnarray}
&&F_{\alpha\beta;\gamma}+F_{\gamma\alpha;\beta}+F_{\beta\gamma;\alpha}=0, \\
&&J^{\mu}=F^{\mu\nu}_{;\nu}=0,
\end{eqnarray}
where  $F_{\mu\nu}=A_{\nu,\mu}-A_{\mu,\nu}$ is the electromagnetic
field tensor. However, the potential (12) given by the
coefficients (14) in the Kerr-Newman metric (including the RN one)
does not exactly satisfy Eq. (16) because the electric charge
density $J^t$ and the current density $J^{\varphi}$ are
nonvanishing for the charged, nonvacuum  black hole metric. This
thing also occurs in the case of nonvacuum black holes under
theories of modified gravity. Extra sources are included in the
theories of modified gravity, therefore, the coefficients (12) or
(13) based on the vacuum black holes are not suitable for the
nonvacuum ones in general. It is necessary to give extensions to
Wald's coefficients (12) or (13) for black holes in
Ho\v{r}ava-Lifshitz gravity [74] and in a braneworld [75].  If the
coefficients $C_{t}$ and $C_{\varphi}$ are assumed to be functions
of the coordinates $r$ and $\theta$, the potential may be solved
from the source-less Maxwell field equations (15) and (16).
Following this idea, Azreg-A\"{i}nou modified the coefficients
(13) as
\begin{equation}
C_{t}=aB+c_t, ~~~~ C_{\varphi}=\frac{B}{2}+c_{\varphi},
\end{equation}
where $c_t$ and $c_{\varphi}$ are functions of the coordinates $r$ and $\theta$ and the parameters
$a$ and $B$. In the
G\"{u}rses-G\"{u}rsey metric [76],
Eq. (12) with Eq. (17) can satisfy Eq. (15) and two
equations $J^{r}=J^{\theta}=0$ of Eq. (16).
When the black hole is nonrotating and spherically symmetrical,
$c_t$ and $c_{\varphi}$ do not depend on $\theta$.  If $g_{tt}=-\left[1-2(f_1r+f_2)/r^2\right]
=-1/g_{rr}$, $c_t$
solved from the equation $J^{t}=0$ is
\begin{eqnarray}
c_{t}=-\frac{\kappa_{1}}{rg_{tt}}-\frac{\kappa_{2}}{g_{tt}},
\end{eqnarray}
where $\kappa_{1}$ and $\kappa_{2}$ are integration constants. In
principle, the two constants can be chosen arbitrarily. However,
they must be $\kappa_{1}=Q$ and $\kappa_{2}=0$ for the black hole
with an electric $Q$, and $\kappa_{1}=\kappa_{2}=0$ for the black
hole that is neutral. Such choices are determined by the Coulomb
potential $-Q/r$, which corresponds to a nonzero covariant
component of the  four-vector electromagnetic potential
$A_t=-(\kappa_{1}/r)-\kappa_{2}$.  The equation $J^{\varphi}=0$
yields
\begin{equation}
c_{\varphi}=\frac{Bf_2}{r^2}.
\end{equation}

In light of the results of Azreg-A\"{i}nou, we can easily write the expression for
the vector potential of electromagnetic field around the hairy black hole (5) as a particular form of the
G\"{u}rses-G\"{u}rsey metric [76].
Here, $Q=0$, $f_1=M$ and $f_2=4\pi \rho_{\mathrm{eff}}$. Considering Eqs. (12), (17)-(19), we have the vector potential
\begin{eqnarray}
A^{\alpha}&=\left(\frac{4 \pi}{r^2}B \rho_{\text{eff}}+\frac{B}{2}\right)\xi^{\alpha}_{\varphi},
\end{eqnarray}
which is equivalent to the following expression
\begin{equation}
A_{\varphi}=\frac{1}{2}B\left(r^2+8\pi\rho _{\text{eff}}\right)\sin ^2\theta.
\end{equation}
This vector potential is an exact solution of the source-less Maxwell field equations (15) and (16).
Note that the vector potential should have been $A_{\varphi}=\frac{1}{2}Br^2\sin^2\theta$ based on the Wald potential,
but such a vector potential satisfies Eq. (15) and the two
equations $J^{r}=J^{\theta}=0$ of Eq. (16) except the other equations $J^{t}=J^{\varphi}=0$ of Eq. (16).

In order to obtain nonzero orthonormal components of the electromagnetic field, we introduce four orthogonal basis vectors
in an observer's reference frame. The observer basis
$\{e_{\hat{t}}, e_{\hat{r}}, e_{\hat{\theta}}, e_{\hat{\varphi}}\}$ is expressed in the coordinate basis
$\{\partial_{ t},\partial_{ r}, \partial_{\theta}, \partial_{\varphi}\}$ as
\begin{eqnarray}
e_{\hat{\mu}}=e_{\hat{\mu}}^{\nu}\partial_{\nu},
\end{eqnarray}
where the transform matrix $e_{\hat{\mu}}$ satisfies the relation $g_{\mu\nu}e_{\hat{\alpha}}^{\mu}e_{\hat{\beta}}^{\nu}=\eta_{\hat{\alpha}\hat{\beta}}$.
$\eta_{\hat{\alpha}\hat{\beta}}$ is the metric of Minkowski spacetime.
For the hairy black hole metric (5), a simple choice of the basis  is given as follows:
\begin{equation}
\begin{matrix}
\\
e_{\hat{\mu}}^{\nu}=\\
\\
\end{matrix}
\begin{bmatrix}
\sqrt{-g^{tt}} & 0 & 0 & 0\\
0 &\sqrt{g^{rr}} & 0& 0\\
0 & 0 &\sqrt{g^{\theta\theta}}& 0\\
0 & 0 & 0&\sqrt{g^{\varphi\varphi}}\\
\end{bmatrix}.
\end{equation}
The 4-velocity of the rest observer $U^{\alpha}$ at the reference frame is
\begin{eqnarray}
&U^{\alpha}=\sqrt{-g^{tt}}(1,0,0,0),
\\
&U_{\alpha}=\sqrt{-g_{tt}}(-1,0,0,0).
\end{eqnarray}
The components of the electric and magnetic fields in the frame read
\begin{eqnarray}
E_{\alpha}&=& F_{\alpha\beta}U^{\beta},
\\
B_{\alpha}&=&\frac{1}{2}\epsilon_{\beta\alpha\mu\nu}F^{\mu\nu}U^{\beta},
\end{eqnarray}
where $\epsilon_{\beta\alpha\mu\nu}$ is the Levi-Civita tensor.
Thus, the nonvanishing orthonormal components of the electromagnetic field measured by the zero angular momentum observer are
\begin{eqnarray}
B_{\hat{r}} &=& B_{r}e_{\hat{r}}^{r}=B(1+\frac{8\pi\rho _{\text{eff}}}{r^{2}})\cos\theta,
\\
B_{\hat{\theta}} &=& B_{\theta}e_{\hat{\theta}}^{\theta}=-B\sin\theta\sqrt{1-\frac{2}{r}-\frac{8\pi\rho _{\text{eff}}}{r^2}}.
\end{eqnarray}
The total magnetic field is $B_{tot}=\sqrt{B_{\hat{r}}^{2}+B_{\hat{\theta}}^{2}}$.
In Fig. 1, we plot $B_{tot}$ varying with the parameter $\rho_\text{{eff}}$
or the radial distance $r$. It is clear that the magnetic field
increases with an increases of the parameter $|\rho _{\text{eff}}|$, but decreases with an increases of  $\theta$.

\section{Motions of charged particles around hairy black holes in external magnetic fields} \label{section3}

Suppose that a  particle with a charge $q$ moving around the hairy black hole with the external magnetic field (21).
The momentum in Eq. (9) is $P_{\mu}=p_{\mu}-qA_{\mu}$. The charged-particle motion
is described by the super-Hamiltonian
\begin{eqnarray}
K &=& \frac{1}{2}g^{\mu\nu}(p_{\mu}-qA_{\mu})(p_{\nu} -qA_{\nu})
\nonumber \\
&=& -\frac{1}{2}\left(1-\frac{2}{r}-\frac{8\pi\rho_{\text{eff}}}{r^2}\right)^{-1}
E^{2} \nonumber \\
&& +\frac{1}{2}\left(1-\frac{2}{r} -\frac{8 \pi\rho_{\text{eff}}}{r^2}\right)p^{2}_{r}
+\frac{1}{2}\frac{p^{2}_{\theta}}{r^2} \nonumber \\
&& +\frac{1}{2r^2\sin^2\theta}\left[L-\frac{b}{2}\left(r^2+8\pi\rho_{\text{eff}}\right)\sin^{2}\theta\right]^{2},
\end{eqnarray}
where $b=qB$. The scale transformations in $K$ are similar to those in $H$. In addition,
$q\rightarrow mq$ and $B\rightarrow B/M$ are used.
Similar to $H$, $K$ still satisfies the constraint
\begin{equation}
K=-\frac{1}{2}.
\end{equation}
However, the external magnetic field leads to the absence of the fourth constant (11).
As a result, the Hamiltonian (30) is nonintegrable. This fact shows that the weak electromagnetic field
can exert an influence on the charged particle dynamics although it gives no contribution to the geometry of spacetime.

\subsection{Explicit symplectic integrators}

Symplectic schemes are naturally viewed as  the most appropriate solvers for long-term integration of
the Hamiltonian system (30) because they preserve the symplectic structure of Hamiltonian dynamics.
Explicit symplectic methods are less than implicit ones at the expense of computational time.
The Hamiltonian is not directly split into two explicitly integrable pieces and then explicit symplectic integrators
become useless. However, they are still variable when the Hamiltonian is split into more than two explicitly integrable parts.
In fact, the construction of explicit symplectic schemes based on the multi-part splitting method has appeared in
recent literature [44-49].

The Hamiltonian (30) is separated in the form
\begin{equation}
K=K_1+K_2+K_3+K_4+K_5,
\end{equation}
where all sub-Hamiltonians are written as follows:
\begin{eqnarray}
K_1 &=& \frac{1}{2r^2\sin^2\theta}\left[L-\frac{b}{2}(r^2+8\pi\rho_{\text{eff}})\sin^{2}\theta\right]^{2} \nonumber
\\ && -\frac{{E}^{2}}{2}\left(1-\frac{2}{r}-\frac{8\pi\rho_\text{eff}}{r^2}\right)^{-1}, \\
K_{2} &=& \frac{1}{2}p^{2}_{r},\\
K_{3} &=& -\frac{1}{r}p^{2}_{r},\\
K_{4} &=& \frac{p^{2}_{\theta}}{2r^2}, \\
K_{5} &=& -\frac{4\pi\rho_\text{eff}}{r^2}p^{2}_{r}.
\end{eqnarray}
It is easy to check that each of the five parts has an analytical solution as
an explicit function of the proper  time $\tau$. Solvers for
the sub-Hamiltonians  $K_1$,
$K_2$, $K_3$, $K_4$, and $K_5$ are termed $\varkappa_1$,
$\varkappa_2$, $\varkappa_3$, $\varkappa_4$ and $\varkappa_5$, respectively.

Setting $h$ as a time step, we have a second-order explicit symplectic integrator
\begin{eqnarray}
S2 (h)= \chi^*(\frac{h}{2})\times \chi(\frac{h}{2}),
\end{eqnarray}
where two first-order solvers are
\begin{eqnarray}
    \chi(h) &=&\varkappa_5(h)\times \varkappa_4(h)\times \varkappa_3(h)\times \varkappa_2(h)\times \varkappa_1(h), \\
    \chi^*(h) &=& \varkappa_1(h)\times \varkappa_2(h)\times \varkappa_3(h)\times \varkappa_4(h)\times \varkappa_5(h).
\end{eqnarray}
Composing three second-order methods, we obtain a fourth-order explicit
symplectic algorithm
\begin{eqnarray}
    S4=S2(\gamma h)\times S2(\delta  h)\times S2(\gamma h),
\end{eqnarray}
where $\gamma =1/(1-\sqrt[3]{2}) $ and $\delta =1-2\gamma$. The construction
is that of Yoshida [77]. By the component of more
first-order operators  $\chi$ and $\chi^*$, an optimized fourth-order  partitioned
Runge-Kutta (PRK) symplectic
algorithm was given in [49] by
\begin{eqnarray}
PRK_64 &=& \chi^* (\alpha _{12} h)\times \chi(\alpha _{11}
h)\times \cdots \nonumber \\ && \times \chi^* (\alpha _2 h) \times
\chi(\alpha _1 h),
\end{eqnarray}
where time coefficients are
\begin{eqnarray}
    \nonumber
    &&\alpha _1=\alpha _{12}= 0.079203696431196,   \\ \nonumber
    &&\alpha _2=\alpha _{11}= 0.130311410182166,   \\ \nonumber
    &&\alpha _3=\alpha _{10}= 0.222861495867608,    \\ \nonumber
    &&\alpha _4=\alpha _9=-0.366713269047426,    \\ \nonumber
    &&\alpha _5=\alpha _8=  0.324648188689706,   \\ \nonumber
    &&\alpha _6=\alpha _7= 0.109688477876750.   \nonumber
\end{eqnarray}

Let us take $h=1$, $E=0.996$, $L=4.6$ and $\rho_\text{{eff}}=0.001$.
The initial conditions are
$r=16$, $\theta=\pi/2$ and $p_r=0$. The initial value $p_{\theta}>0$
is given by Eqs. (29) and (30). The magnetic field parameters are $b=10^{-4}$
for Orbit 1 and $b=10^{-3}$ for Orbit 2. When the integration time reaches
$10^{7}$, the three methods S2, S4 and $PRK_64$ give no secular drifts to
Hamiltonian errors $\Delta K=K+1/2$ for integrations of the two orbits in Fig. 2a, b.
They exhibit an advantage of symplectic methods in the energy conservation.
It is also shown that S4 is
four orders of magnitude better than S2 but two orders of magnitude poorer than  $PRK_64$ in accuracy.
Thus, the algorithm $PRK_64$ is employed in later computations.

\subsection{Chaos detection methods}

The phase space structures of Orbits 1 and 2 in Fig. 2a, b can be described through the Poincar\'{e} map in the two dimensional
$r-p_r$ plane. In fact, these points $(r,p_r)$ obtained from the Poincar\'{e} map are intersections of the
particles' orbits with the surface of section $\theta=\pi/2$ and $p_{\theta}>0$. Orbit 1 is regular and nonchaotic because
the intersection points form a closed curve in Fig. 2c. However, Orbit 2 is chaotic because
the intersection points behave in a random distribution way.  The Poincar\'{e} map method
is useful to classify whether a number of orbits in a conservative system with four-dimensional phase space are regular
or chaotic.

The maximum  Lyapunov exponent (mLE) is also a common tool to distinguish between regular motions and irregular ones.
It is often used to quantify the rate of divergence between nearby orbits.
If a bounded orbit has a positive Lyapunov exponent, it is chaotic; If the  mLE of a bounded orbit
vanishes, this orbit is ordered. It takes a very long time to calculate the mLE in a weakly chaotic orbit.
The  fast Lyapunov indicator
(FLI) of Froeschl\'{e} et al. [54] is
a quicker  chaos indicator to investigate such a weak chaotic property. It was developed as a relativistic
invariant form of two nearby orbits [55] from a modified version of the relativistic
invariant Lyapunov exponent with two nearby orbits [56]. The linear growth of FLI with time shows the regularity of Orbit 1 in Fig. 2d,
whereas the exponential growth of FLI indicates the chaoticity of Orbit 2.

Compared with the techniques of Poincar\'{e} map,  mLE and FLI,
the recurrence analysis method [23,60-63] is rarely used to detect chaos from order in relativistic astrophysics.
The method relates to the description of recurrence plots (RPs),
which measure the recurrences of an orbit into the vicinity of previously reached phase-space points
in terms of the recurrence quantification analysis. The method is described as follows.
For a given phase-space variable $\textbf{x}(\tau)$ of an orbit at time $\tau$ in a dynamical system,
the recurrence matrix is defined as
\begin{equation}
\label{rpdef}
\mathbf{R}_{ij}(\varepsilon)=\Theta(\varepsilon-||\textbf{x}(i)-\textbf{x}(j)||) ~~
(i,j=1,...,N).
\end{equation}
Here, $\Theta$ is the Heaviside function: $\Theta(\vartheta)=0$ for $\vartheta<0$ and $\Theta(\vartheta)=1$ for $\vartheta\geq0$.
$\varepsilon$ stands for a pre-defined threshold parameter.
$N$ represents the sampling number. When $T$ is the total integration time, one of the sampling number
corresponds to the time $T/N$.
In this sense, $i$ denotes the time $\tau_i=iT/N$. $||\;.\;||$ is the Euclidean norm $L^2$.
A visual plot of the recurrence matrix $\mathbf{R}_{ij}$ is made of points $(i,j)$, which
correspond to the binary values 0 and 1. A black dot represents the pair $(i,j)$ for $\mathbf{R}_{ij}=1$ and a
white dot denotes the pair $(i,j)$ for $\mathbf{R}_{ij}=0$. Because $\mathbf{R}_{ij}=\mathbf{R}_{ji}$,
the visual is symmetric with respect to the main diagonal, i.e.
the line of identity $j=i$. The visual behaviour of points $(i,j)$ about the presence or absence
of diagonal structures  can contain wealth of dynamical information.
If there are many diagonal lines  parallel to the main diagonal, the considered orbit
is regular. If there are short, disrupted diagonal features or no diagonal lines  parallel to the main diagonal, the motion is chaotic.
In a word, the RP behaves in regular diagonal structures for the nonchaotic case,
but has more complicated, irregular structures for the chaotic case.
The regular or chaotic dynamics of an orbit can be characterized via the visual behaviour of points $(i,j)$
corresponding to the binary values 0 and 1 in a two-dimensional plane.

We take $N=1000$, $T=10^{7}$ and $\varepsilon=k\sigma$, where $\sigma$ is the standard mean deviation of the given data set and $k$ is a proportionality constant [61].
Although  $\textbf{x}(\tau)$ is taken as the  phase-space variables $(r,\theta,p_r,p_{\theta})$, any one of the  phase-space variables
is admissible. For example, $r$ is given to $\textbf{x}(\tau)$. We draw the RPs of Orbits 1 and 2 in Fig. 2e, f.
The presence of a number of  diagonal lines  parallel to the main diagonal
shows the regularity of Orbit 1.  The absence of diagonal lines  parallel to the main diagonal
determines the chaoticity of Orbit 2. The RP method is an efficient tool to identify the dynamical features of Orbits 1 and 2,
as the techniques of Poincar\'{e} surfaces of section and FLIs are.

The visual plot of $\mathbf{R}_{ij}$ for the quasiperiodic orbit 1
is shown in Fig. 2e. What is a visual for a periodic orbit? In
order to answer this question, we choose some circular orbits of
particles in the system (9). Based on Eq. (11) with
$P_r=P_{\theta}=0$, the effective potential for particles moving
at the plane $\theta$ is
\begin{eqnarray}
V_{eff} &=& E^2=\left(1+\frac{C_k}{r^2}\right)\left(1-\frac{2}{r}-\frac{8\pi\rho_{\text{eff}}}{r^2}\right), \\
C_k &=& L^2 \csc^2\theta.
\end{eqnarray}
When $L=4$ and $\rho_{\text{eff}}=0.001$ are given, the effective
potentials at the three planes $\theta=\pi/4,~\pi/3,~\pi/2$ are
shown in Fig. 3a. The conditions for stable circular orbits at the
planes $\theta$ are $dV_{eff}/dr=0$ and $d^2V_{eff}/dr^2\geq0$.
The stable circular orbits in Fig. 3b correspond to their
parameters and radii as follows: $E=0.98$ and $r=28.62$ for the
circular orbit 1 at the plane $\theta=\pi/4$; $E=0.97$ and
$r=17.68$ for the circular orbit 2 at the plane $\theta=\pi/3$;
$E=0.96$ and $r=11.95$ for the circular orbit 3 at the plane
$\theta=\pi/2$. When the circular orbit conditions and the
algorithm $PRK_64$ are applied to the Hamiltonian (30) with $b=0$,
we plot the visuals of $\mathbf{R}_{ij}$ for these stable circular
orbits. The visual for the circular orbit 1 has diagonal lines
parallel to the main diagonal and numerous square lattices in Fig.
3c. The visual for the circular orbit 3 in Fig. 3d looks like that
for the quasiperiodic orbit 1 in Fig. 2e and has a series of
diagonal lines parallel to the main diagonal. This fact shows that
the RPs of periodic orbits and those of quasiperiodic orbit are not strictly
distinguishable.

\subsection{The effect of varying one parameter on a transition from regular to chaotic
regime}

We use the method of
FLIs to study the effect of varying one parameter
on the transition from regular to chaotic regime in the
Hamiltonian (30) with $b\neq0$. For comparison, the methods of Poincar\'{e} surfaces of section and RPs are
also employed.

Taking the parameters $L=4.6$,  $b=0.0001$, $\rho_\text{{eff}}=0.0001$ and the initial conditions
$r=16$, $\theta=\pi/2$, we estimate the FLIs of 30 trajectories with the energy
running over the interval $E\in [0.9970,0.9999]$. The FLIs in Fig. 4a show the occurrence of abrupt transitions to chaos
at $E=0.9977$ and $E=0.9992$. Each of the FLIs is obtained after the integration time arrives at
$\tau=2\times10^{6}$. All FLIs larger than (or equal to) 15 correspond to the onset of chaos, but those
less than this value indicate the regular dynamics. The FLIs at $E\leq0.9976$ indicate the regular dynamics.
The trajectory for the energy $E=0.9976$ is a torus on the Poincar\'{e} surface of section
in Fig. 4b, and its regularity is also confirmed through the RP with diagonal lines
parallel to the main diagonal in Fig. 4f. When $E=0.9977$, the weak chaoticity is shown by the methods of Poincar\'{e} map and FLI in Fig.
4a, b.  The RP in Fig. 4e  seems to have diagonal line structures that are slightly disrupted. The regular dynamics
exists for $0.9978\leq E\leq 0.9992$, but the chaotic dynamics
does for $E\geq0.9993$. Especially for $E=0.9996$ and $E=0.9998$, the chaotic behaviours are described in Fig. 4b,
and are also shown by the RPs in Fig. 4c, d. The RPs have complex large-scale torus structures unlike diagonal line structures.
Fig. 4 shows that the degree of chaos increases when the energy increases in the interval $0.9992\leq E\leq 0.9996$.
In fact, the degree of chaos is strengthened with the energy increasing from a global phase-space structure under appropriate conditions.

Now, we consider the effect of the angular momentum on the chaotic motion. When the parameters $E=0.996$,  $b=0.0006$, $\rho_\text{{eff}}=0.0001$ and the initial conditions
$r=16$, $\theta=\pi/2$ are given, $L$ ranges from 4.0 to 4.2.
The FLIs in Fig. 5a show that  the transition from chaotic to regular
regime occurs at the angular momentum $L=4.07$. It is seen clearly  that the extent of chaos decreases
with the angular momentum increasing. The Poincar\'{e} maps in Fig. 5b and the RPs without complete diagonal line structures
in Fig. 5c, d give chaotic dynamical information for $L=4.03$, 4.05. Although the diagonal lines seem to be present for $L=4.05$, they are shorter and then
indicate the weak chaoticity. The trajectories are ordered for $L=4.07$, 4.10, as shown via
the Poincar\'{e} maps in Fig. 5b and the RPs with diagonal line structures in Fig. 5e, f.

Then, let the magnetic field parameter $b$ be varied in the
interval $b\in[-0.001,0.001]$, where the other parameters are  $E
= 0.996$, $L = 4.4$, $\rho_\text{{eff}}== 0.0001$ and the initial
separation is  $r = 10$. The FLIs in Fig. 6a have abrupt changes
at $b=-0.0008$, 0.0003, 0.0006, 0.0008. As claimed below Eq. (31),
the external magnetic field destroys the existence of a fourth
constant in the Hamiltonian system (30), and thus, it should be
responsible for chaotic dynamics of charged particles. Even the
small values of the magnetic parameter such as $b=-0.0008$, 0.0008
can exert strong chaotic effects on the trajectories of charged
particles in Fig. 6b. The chaotic behaviours at $b=-0.0008$,
0.0008 are also described via the RPs in Fig. 6c, d.

Finally, we investigate the dependence of regular and chaotic dynamics on varying the hairy parameter $\rho_\text{{eff}}$. In various simulations,
the hairy parameter is constrained in the interval $\rho_\text{{eff}}\in [-0.1, 0.1]$, which comes from the
limits of $\rho_\text{{eff}}$ based on the Event Horizon Telescope (EHT) observations of Sagittarius A$^*$
(Sgr A$^*$) [67]. The other parameters are $E=0.996$, $L=4.6$, $b=0.00088$
and the initial conditions are $r=16$, $\theta=\pi/2$.
Chaos exists for $\rho_\text{{eff}}\geq0.005$ in Fig. 7a about the FLIs depending on $\rho_\text{{eff}}$.
The chaoticness of charged-particle motions for $\rho_\text{{eff}}=0.005$, 0.01 is checked
by the Poincar\'{e} maps in Fig. 7b and the RPs without diagonal line structures in Fig. 7c, d.
If the parameters and one of the initial conditions are altered as $E=0.999$, $b=0.00001$
and $r=9$, the FLIs in Fig. 8a correspond to chaos for $-0.001\leq \rho_\text{{eff}}\leq-0.003$ and $\rho_\text{{eff}}=0,~0.005,~0.006$.
The regular dynamics is also shown for
$\rho_\text{{eff}}=-0.002, ~-0.001$, $0.001\leq \rho_\text{{eff}}\leq0.004$,
and $0.007\leq \rho_\text{{eff}}\leq 0.01$.
The chaoticity for $\rho_\text{{eff}}=0$ and the regularity for $E=0.01$
can be observed from the RPs in Fig. 8c, d. When the parameters and one of the initial conditions become $E=0.996$, $L=4.1$, $b=-0.00008$
and $r=6$, there is
an abrupt transition to chaos when $\rho_\text{{eff}}$ exceeds -0.006, as is seen from the FLIs in Fig. 9a. Fig. 9b-d describe
the regular dynamics at $\rho_\text{{eff}}=-0.01$ and the chaotic dynamics at  $\rho_\text{{eff}}=0.005$.
Because of different choices of the initial conditions and other parameters in Figs. 7a, 8a and 9a, there is no universal rule for the dependence
of chaotic dynamics on  the hairy parameter $\rho_\text{{eff}}$.

The above demonstrations completely support the results of [23]. The results are summarized here. The visuals of RPs for periodic or quasiperiodic orbits
exhibit typical regular structures on the diagonal
lines parallel to the main diagonal, as shown in Figs. 2e, 3c, d, 4f, 5e,f, 8d and 9d.
The visuals of RPs for weakly chaotic orbits still have the contour of diagonal line structures, but the structures are short or slightly disrupted. Thus,
no strict diagonal line structures are present, as shown in
Figs. 4e and 5c, d.  There are no diagonal line structures  or complex, irregular large-scale
torus structures unlike diagonal line structures in the visuals of RPs for the existence of strong chaos, as shown in Figs. 2f, 4c, d, 6c, d,
7c, d, 8c and 9c.

\section{Conclusions} \label{section4}

The Horndeski gravity is a theory of modified gravity based on a
very general scalar-tensor theory. A hairy black hole solution in
the Horndeski gravity is  spherically symmetric and asymptotically
flat. It is the RN-like black hole solution for a negative value
of the hairy parameter. An asymptotically uniform magnetic field
in the vicinity of the hairy black hole is so weak that it has a
negligible effect on the spacetime background but exerts a large
influence on the motion of charged test particles. The vector
potential of electromagnetic field in the context of modified
gravity must be a modified version of the Wald potential derived
from the vacuum background. Such a modified vector potential can
strictly satisfy the source-less Maxwell equations.

Explicit symplectic integrators exhibit excellent long-term behaviour in simulating the motion of charged particles around the hairy black hole
immersed in the external magnetic field.
Chaos indicators such as the methods of Poincar\'{e} surfaces of section, FLIs
and RPs are used to investigate the regular and chaotic dynamics of charged particles. The RP method is the recurrence quantification analysis method,
which measures the recurrences of an orbit into the vicinity of previously reached phase-space points.
A visual plot of the recurrence matrix with or without diagonal structures parallel to the main diagonal can contain wealth of dynamical information.
The presence of diagonal structures  means the regular dynamics,
but the absence of diagonal structures or the existence of short, disrupted diagonal features
shows the chaotic dynamics. The RP method is efficient to detect chaos from order, as the methods of Poincar\'{e} surfaces of section and FLIs are.

\section*{Acknowledgments}

The authors are also very grateful to a referee for valuable comments and suggestions.
Author Cao also thanks Dr. Menghe Wu for useful discussions on the
electromagnetic field around the hairy black hole in Horndeski gravity.
This research has been supported by the National Natural Science
Foundation of China (Grant No. 11973020).

\textbf{Data Availability Statement}: This manuscript has no
associated data or the data will not be deposited. [Author's
comment: All of the data are shown as the figures and formula. No
other associated data.]

\textbf{Code Availability Statement} Code/software will be made
available on reasonable request. [Author's comment: The
code/software generated during and/or analysed during the current
study is available from the first author W. Cao on reasonable
request.]

\begin{figure*}[htbp]
\center{
\includegraphics[scale=0.5]{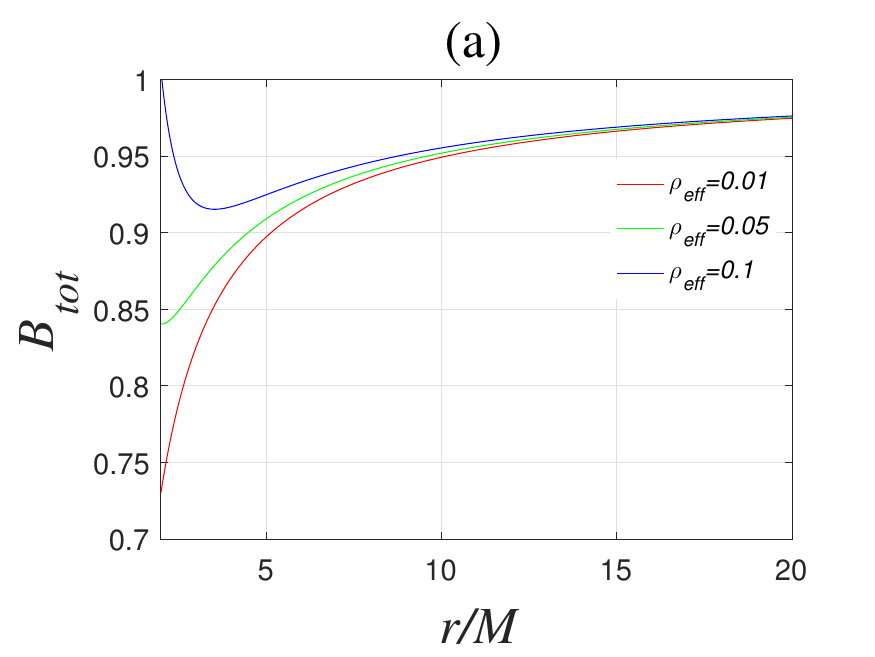}
\includegraphics[scale=0.5]{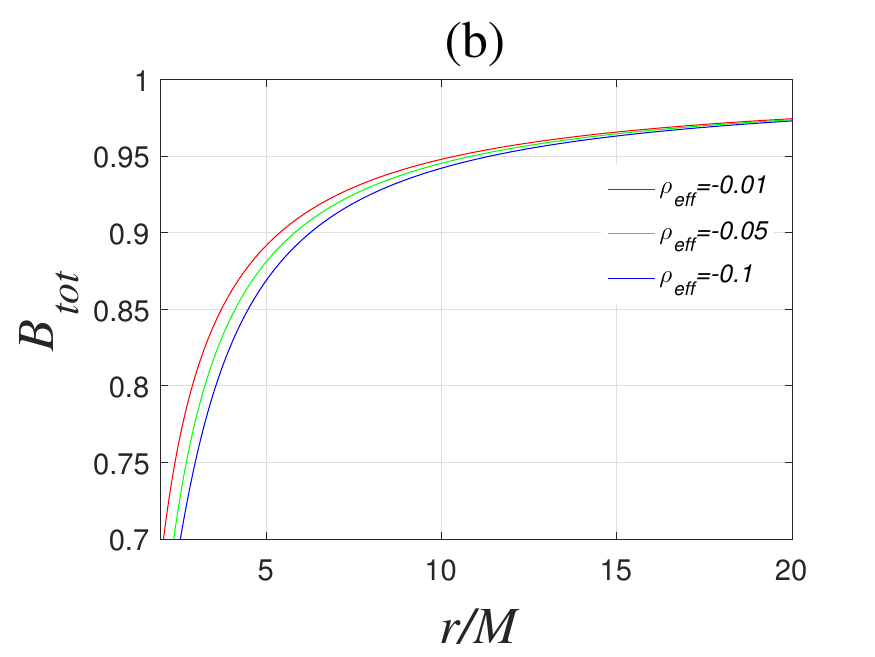}
\includegraphics[scale=0.5]{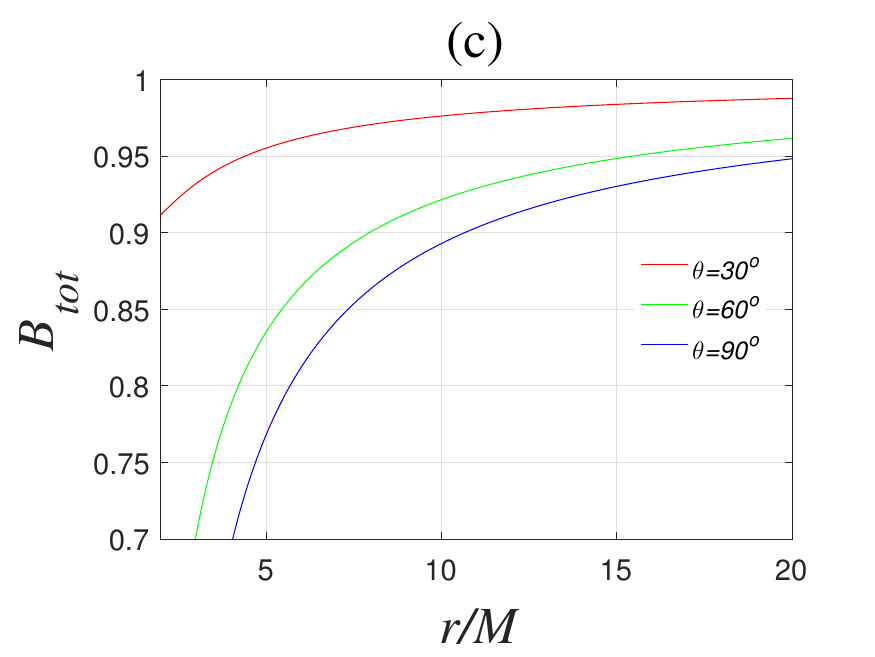}
\includegraphics[scale=0.5]{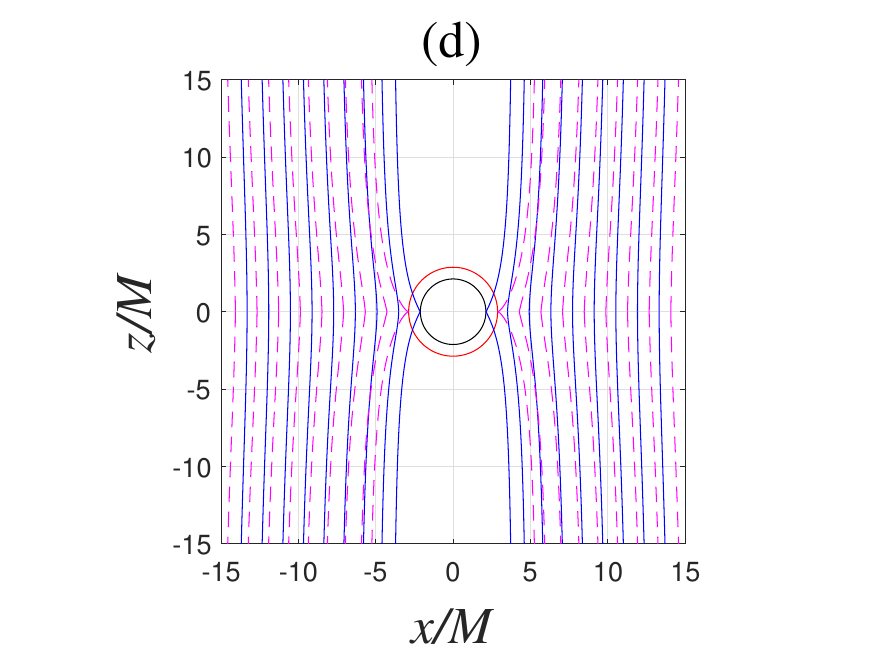}
\caption{Magnetic configurations of the hairy black hole. (a): The
total magnetic field $B_{tot}$ with $B=1$ and $\theta=\pi/4$
for three positive values of the hairy parameter
$\rho_\text{{eff}}$. (b): The total magnetic field $B_{tot}$ for
three negative values of the hairy parameter $\rho_\text{{eff}}$.
(c): The total magnetic field $B_{tot}$ with $B=1$ and
$\rho_\text{{eff}}=0.01$ for different observational angles
$\theta$. (d): Magnetic field lines  in the vicinity of the hairy
black hole in the $x-z$ plane, where $B=1$,
$\rho_\text{{eff}}=0.001$ for blue lines, and
$\rho_\text{{eff}}=0.01$ for pink lines. }
 \label{Fig1}}
\end{figure*}

\begin{figure*}[htbp]
\center{
\includegraphics[scale=0.3]{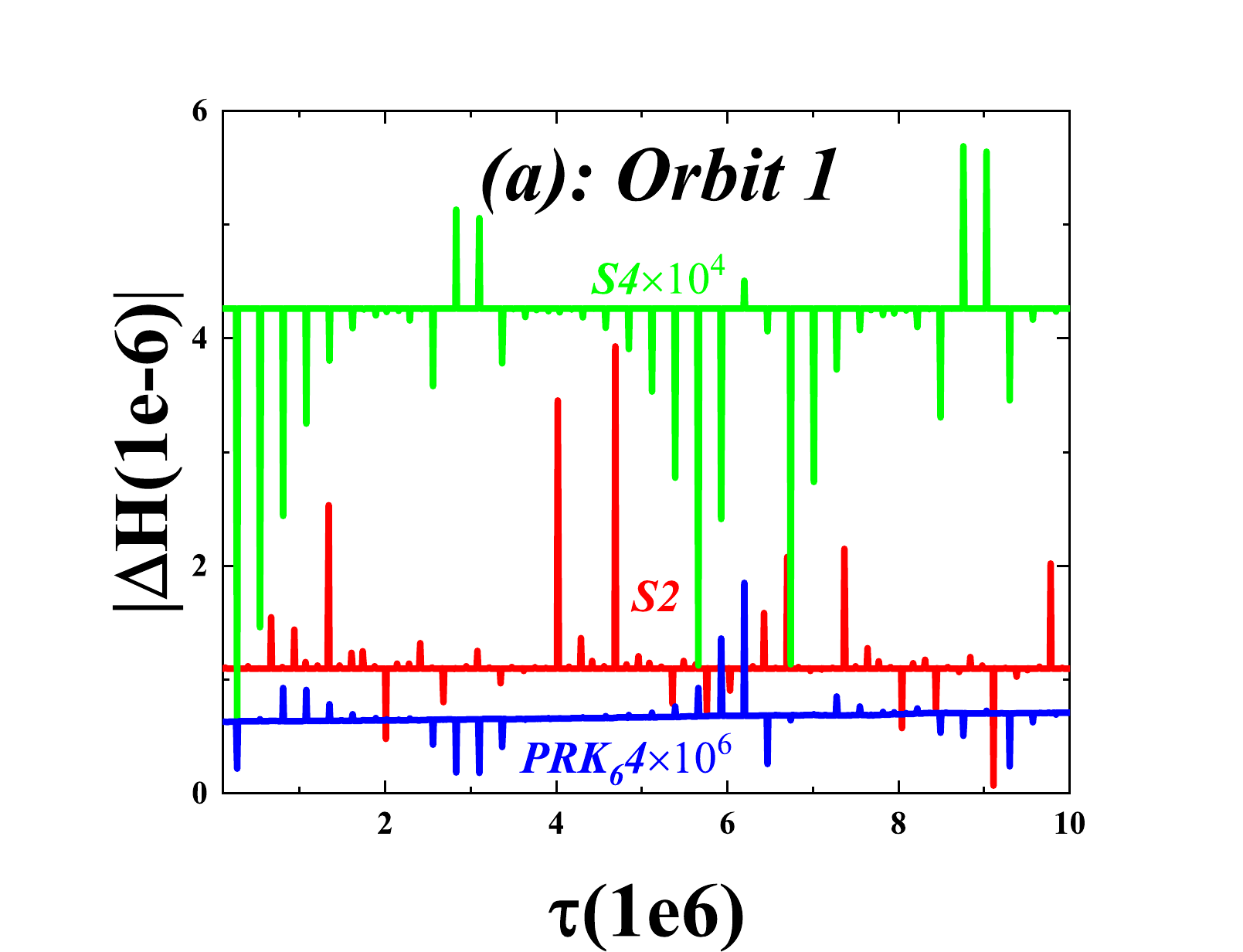}
\includegraphics[scale=0.3]{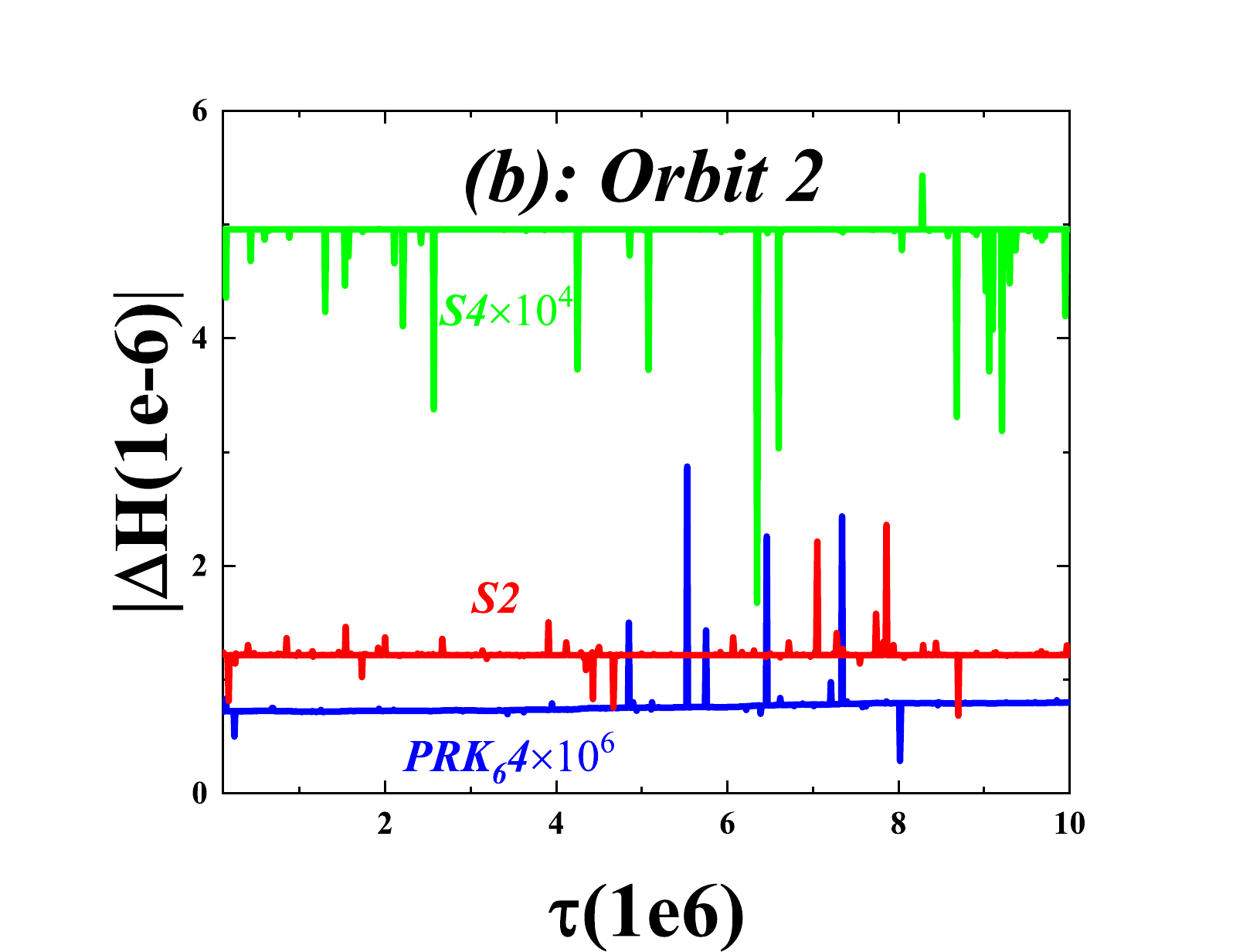}
\includegraphics[scale=0.3]{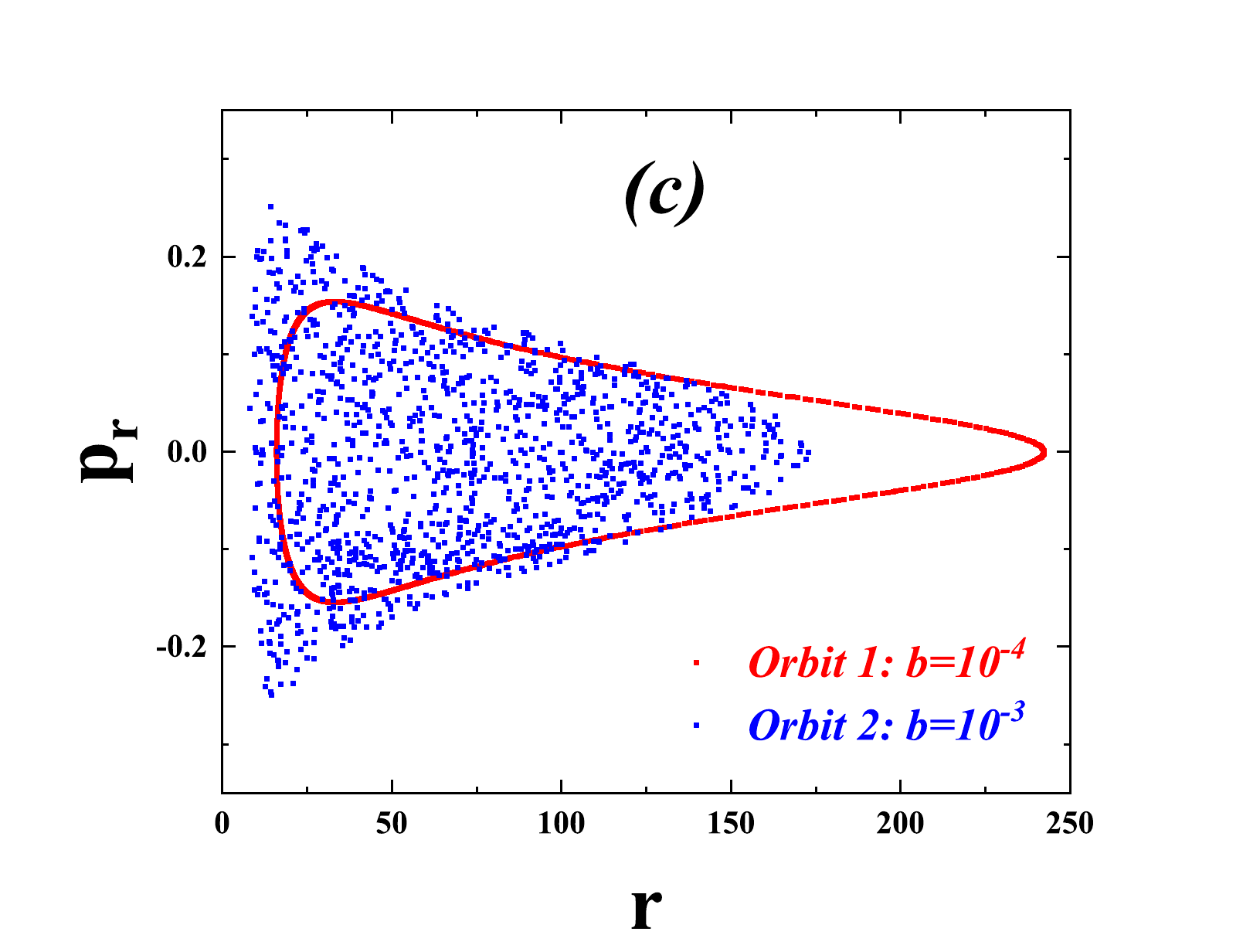}
\includegraphics[scale=0.3]{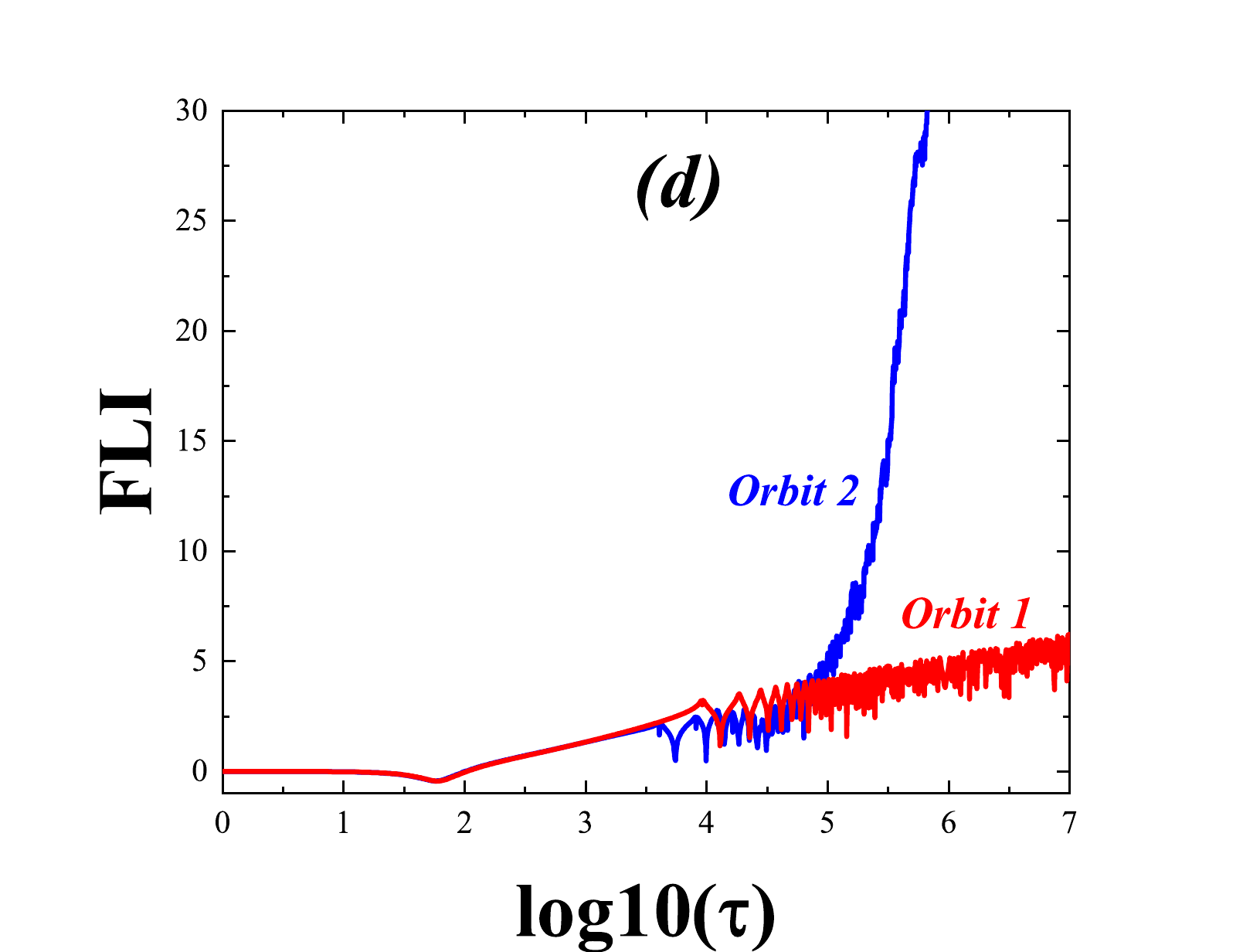}
\includegraphics[scale=0.5]{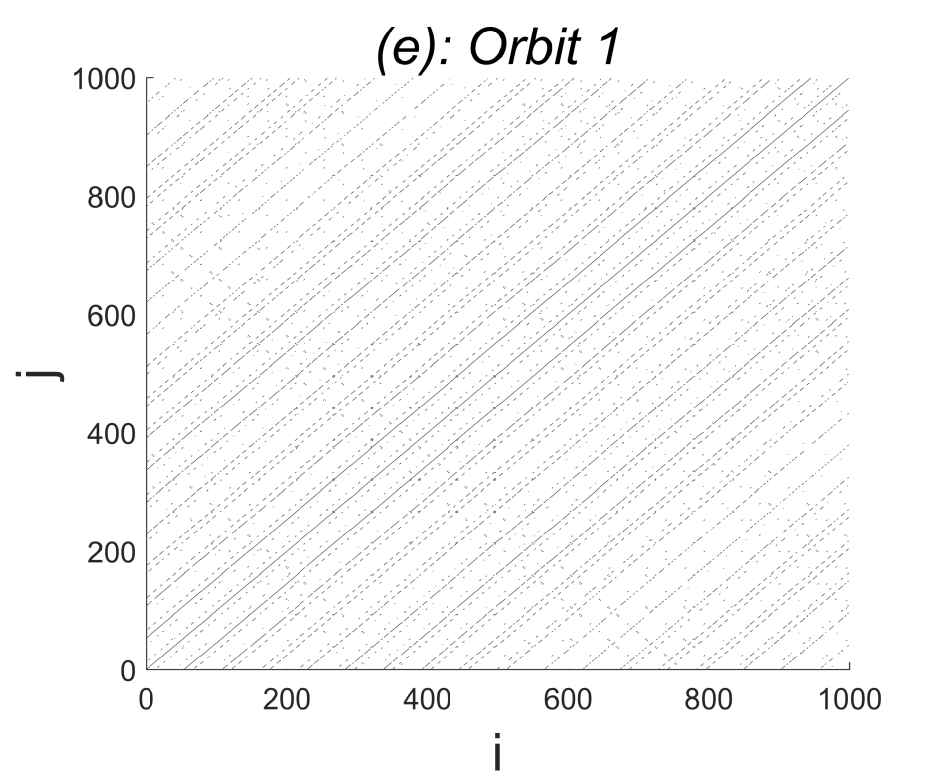}
\includegraphics[scale=0.5]{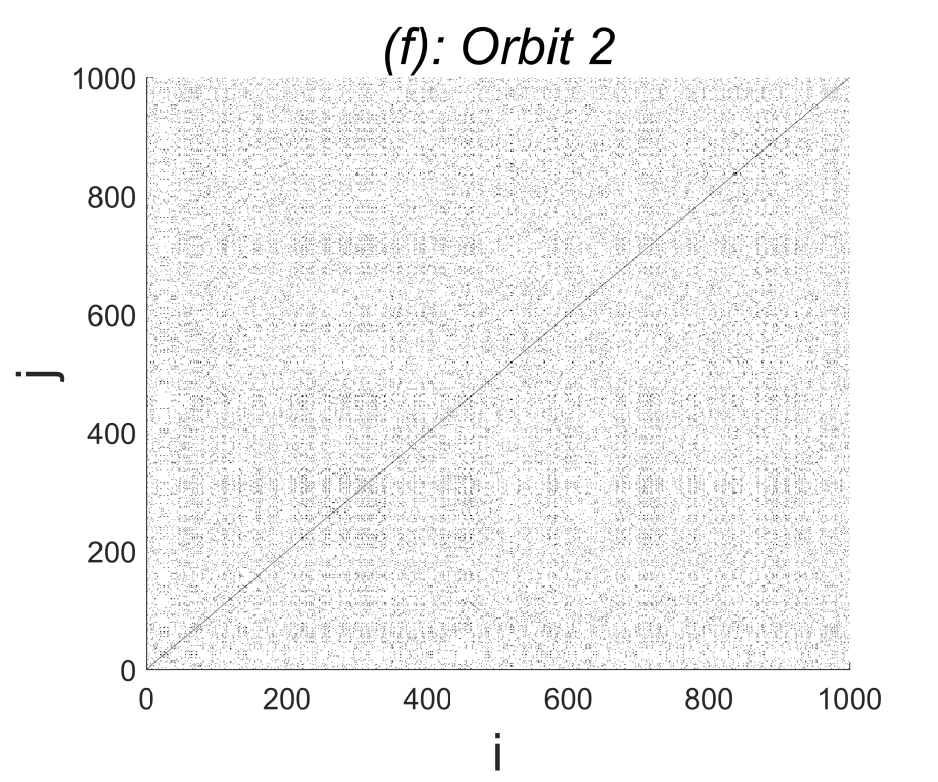}
\caption{(a) and (b): Hamiltonian errors $\Delta K=K+1/2$ for
several explicit symplectic algorithms integrating Orbits 1
and 2, which have common parameters $E=0.996$, $L=4.6$,
$\rho_\text{{eff}}=0.001$ and initial conditions $r=16$,
$\theta=\pi/2$, but different magnetic parameters $b$. (c):
Poincar\'{e} sections of the two orbits. (d): FLIs of the two
orbits. (e) and (f): RPs of the two orbits. Here, $i$ corresponds to the time $\tau_i=i\times10000$.
Many diagonal lines parallel to the main diagonal $j=i$ show
the regular dynamics of Orbit 1, whereas no diagonal lines parallel to the main diagonal describe the chaotic dynamics
of Orbit 2. That is, the dynamical features of Orbits 1 and 2 described by the RPs are consistent with those given by the
techniques of Poincar\'{e} sections and FLIs.}
 \label{Fig2}}
\end{figure*}

\begin{figure*}[htbp]
\center{
\includegraphics[scale=0.5]{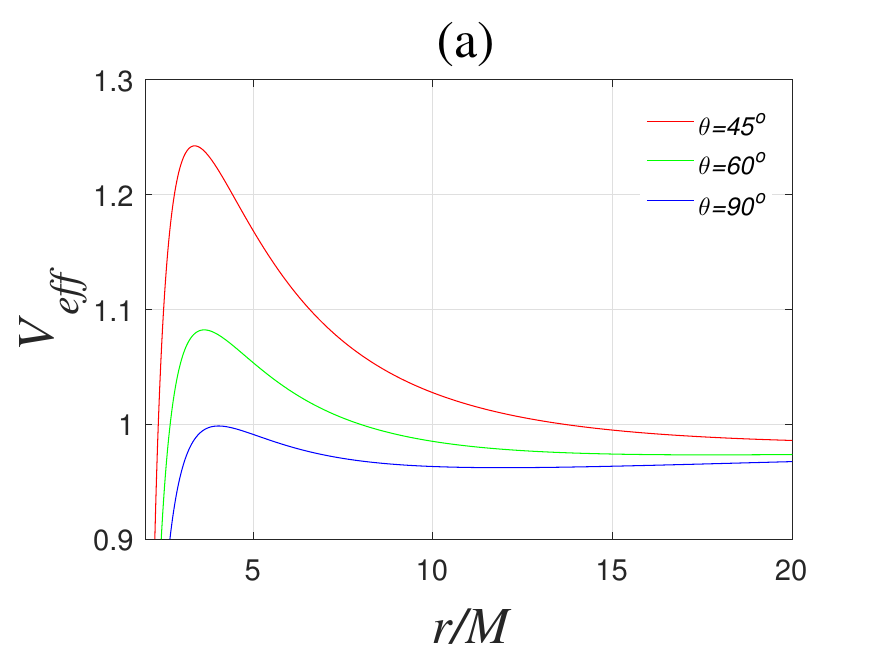}
\includegraphics[scale=0.3]{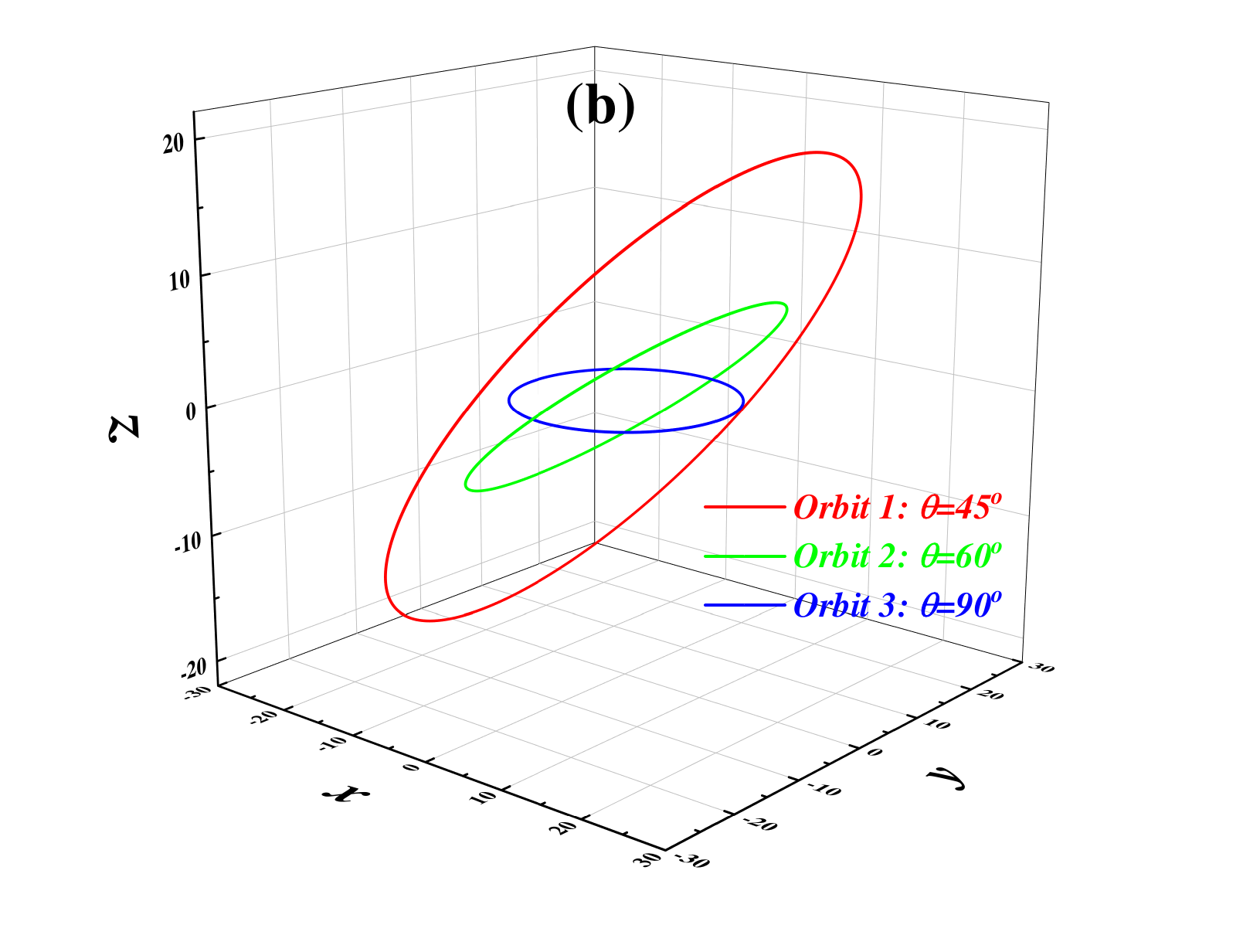}
\includegraphics[scale=0.5]{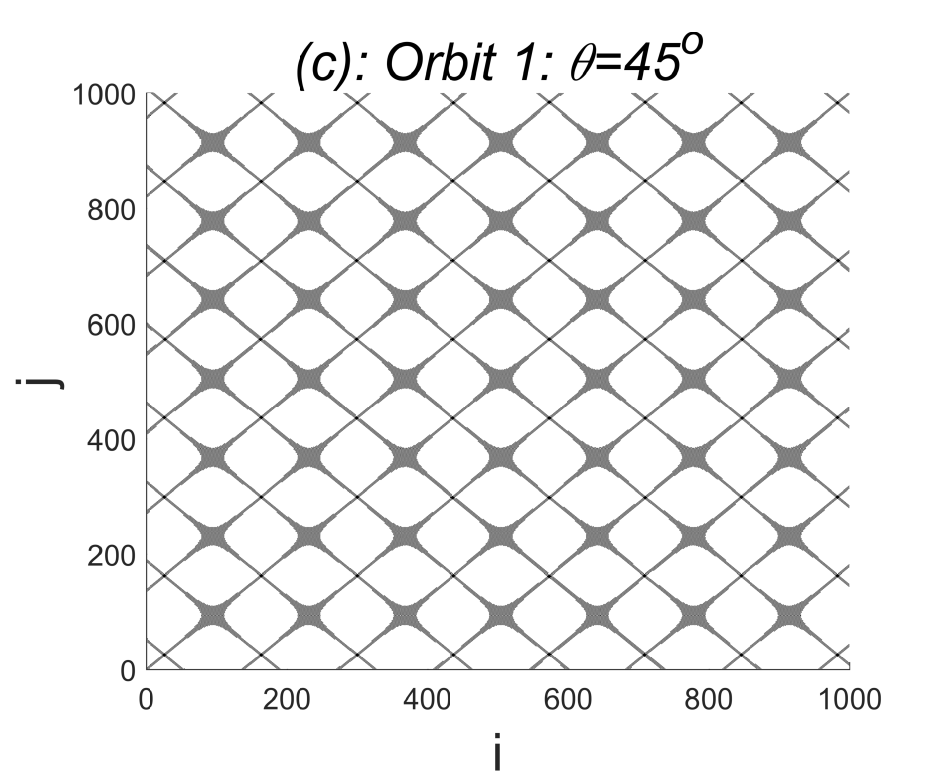}
\includegraphics[scale=0.5]{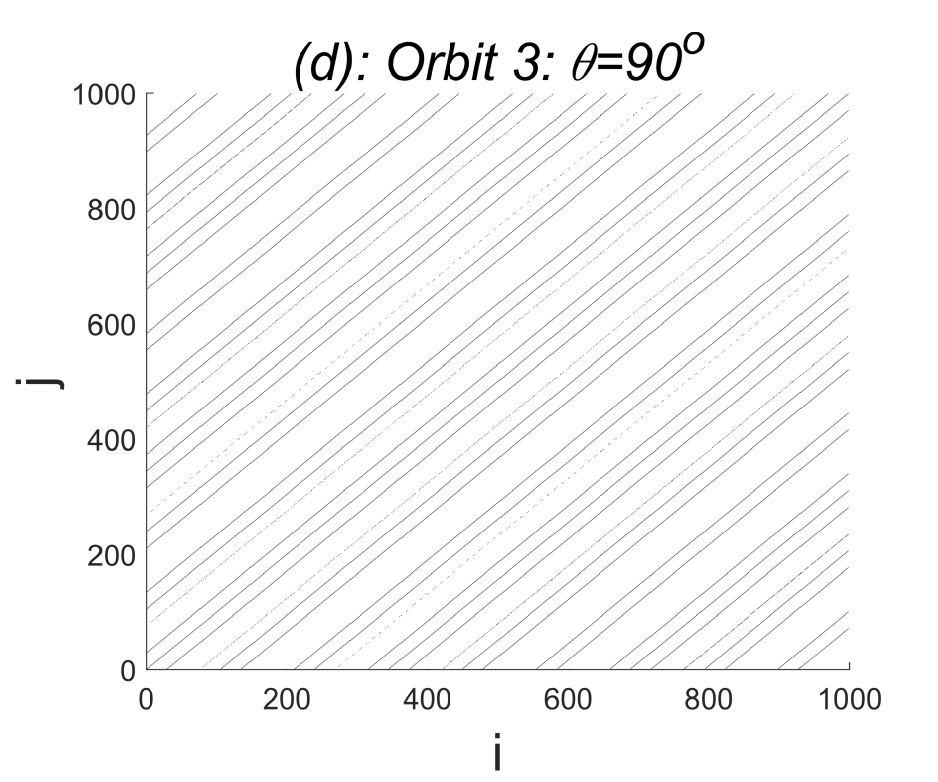}
\caption{(a) Effective potentials for uncharged test particles
moving at three different planes, where  common parameters are
$L=4$ and $\rho_\text{{eff}}=0.001$. (b) Stable circular orbits at
the three different planes. (c) RP of the circular orbit  at the
plane $\theta=\pi/4$ corresponding to $E=0.98$ and the radius
$r=28.62$. (d) RP of the circular orbit  at the plane
$\theta=\pi/2$ corresponding to $E=0.96$ and the radius
$r=11.95$.} \label{Fig3}}
\end{figure*}

\begin{figure*}[htbp]
\center{
\includegraphics[scale=0.3]{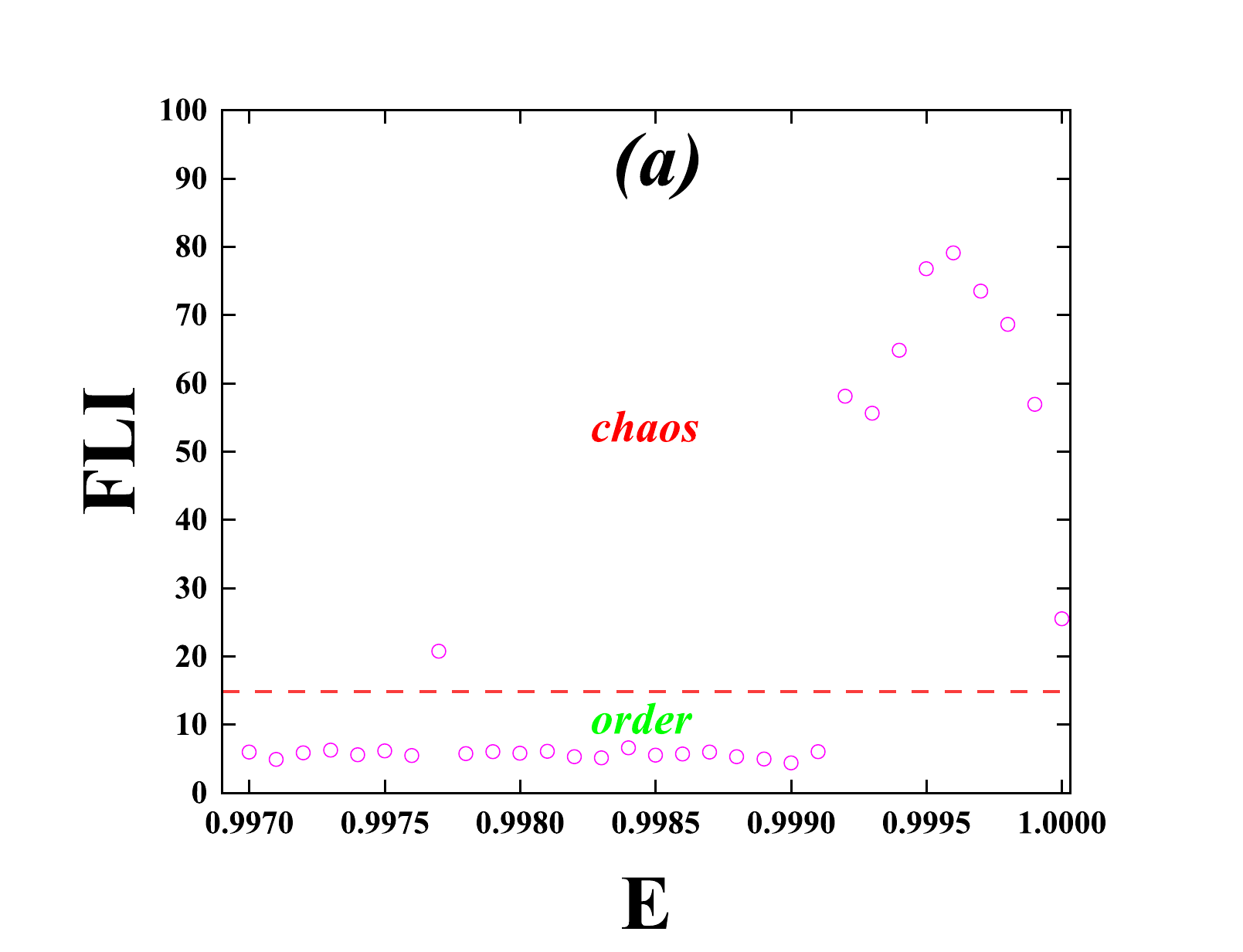}
\includegraphics[scale=0.3]{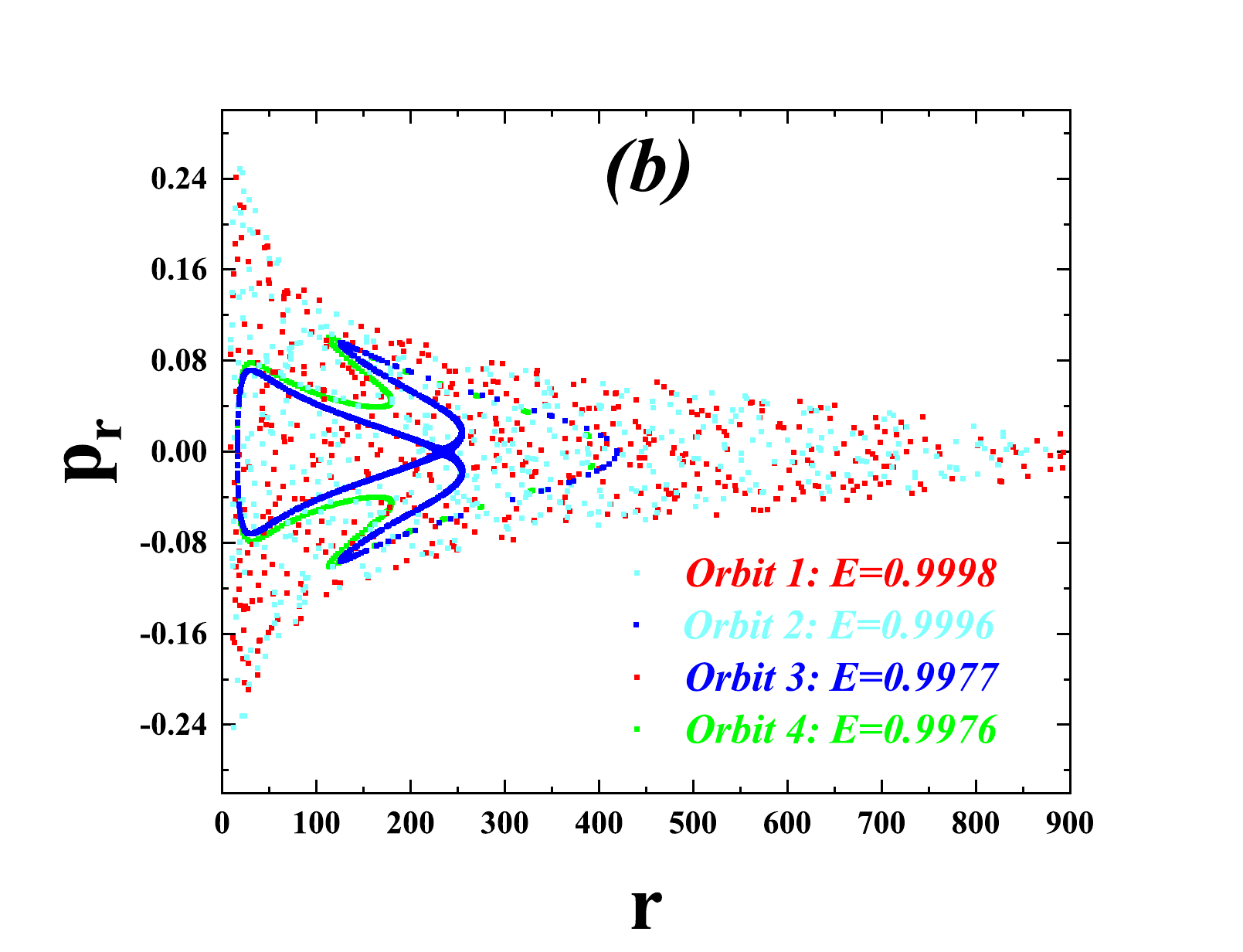}
\includegraphics[scale=0.5]{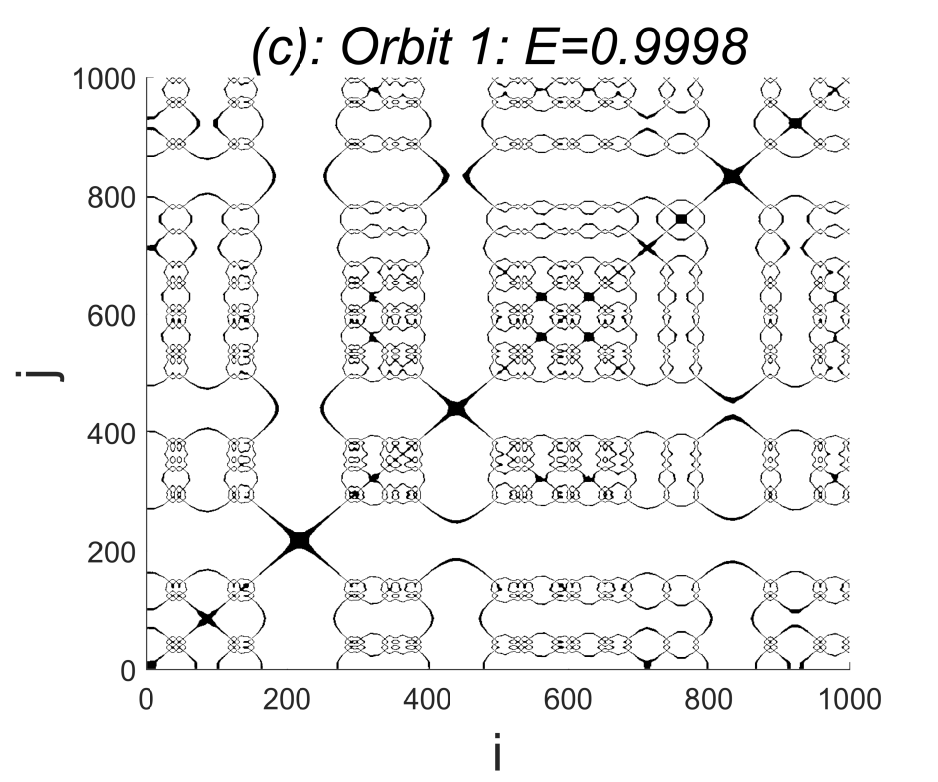}
\includegraphics[scale=0.5]{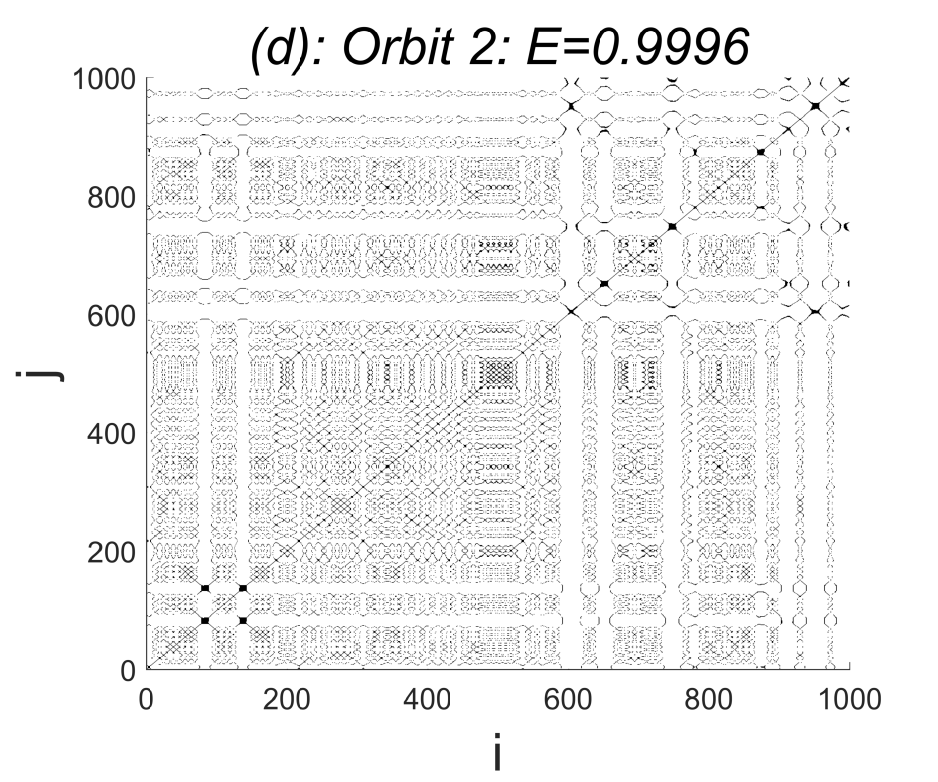}
\includegraphics[scale=0.5]{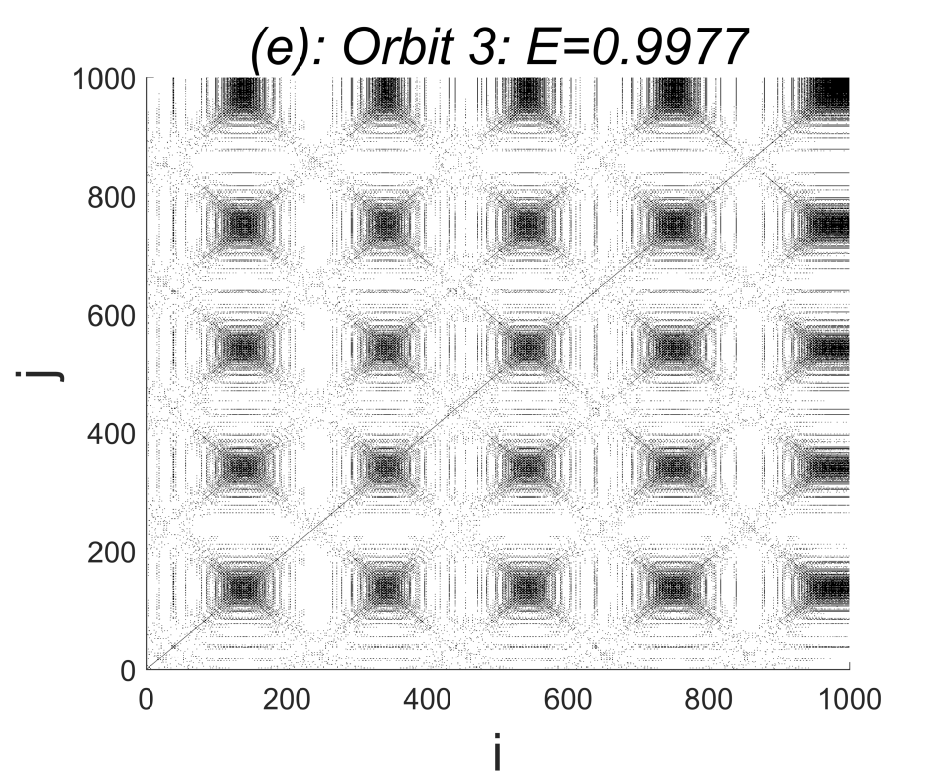}
\includegraphics[scale=0.5]{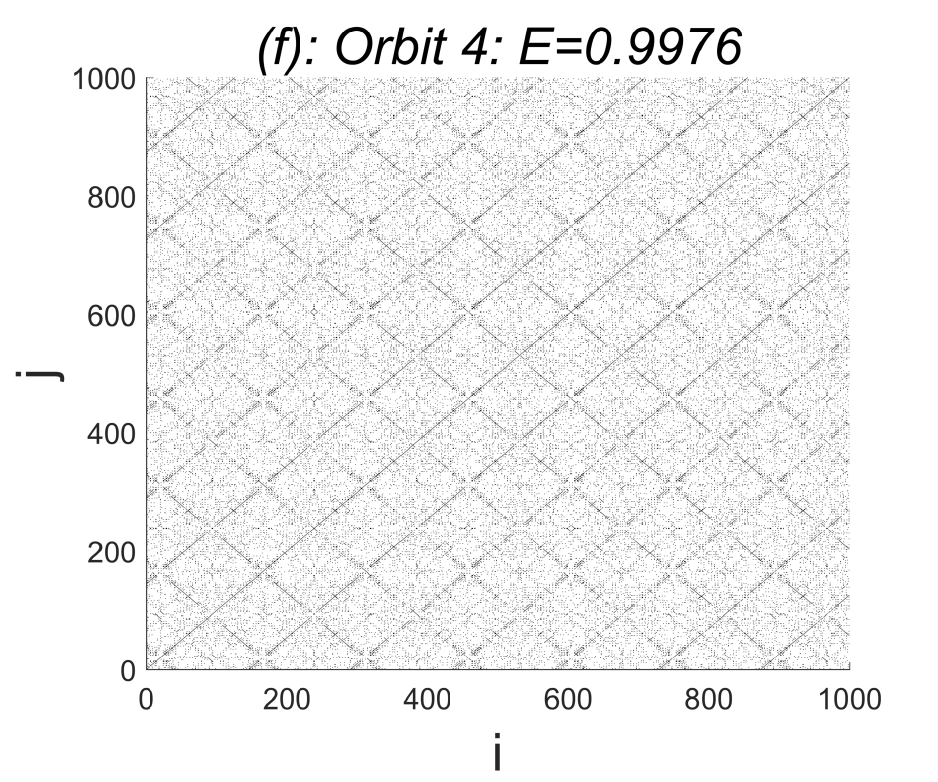}
\caption{(a) FLI describing a dynamical transition to chaos with
the energy $E$ increasing, where the initial conditions are
$r=16$, $\theta=\pi/2$, and the other parameters are $L=4.6$,
$b=0.0001$, $\rho_\text{{eff}}=0.0001$. Chaos occurs at $E=0.9977$ and $0.9992\leq E<1$. (b) Poincar\'{e} sections
for four values of the energy $E$. (c)-(f): RPs for four values of
the energy $E$. The RPs in (c) and (d) correspond to strong chaos.
The RP in (e) seems to exhibit a symmetrical structure with diagonal lines parallel to the main diagonal, while
corresponds to weak chaos. The RP in (f) shows
the regular dynamics. }
 \label{Fig4}}
\end{figure*}

\begin{figure*}[htbp]
\center{
\includegraphics[scale=0.3]{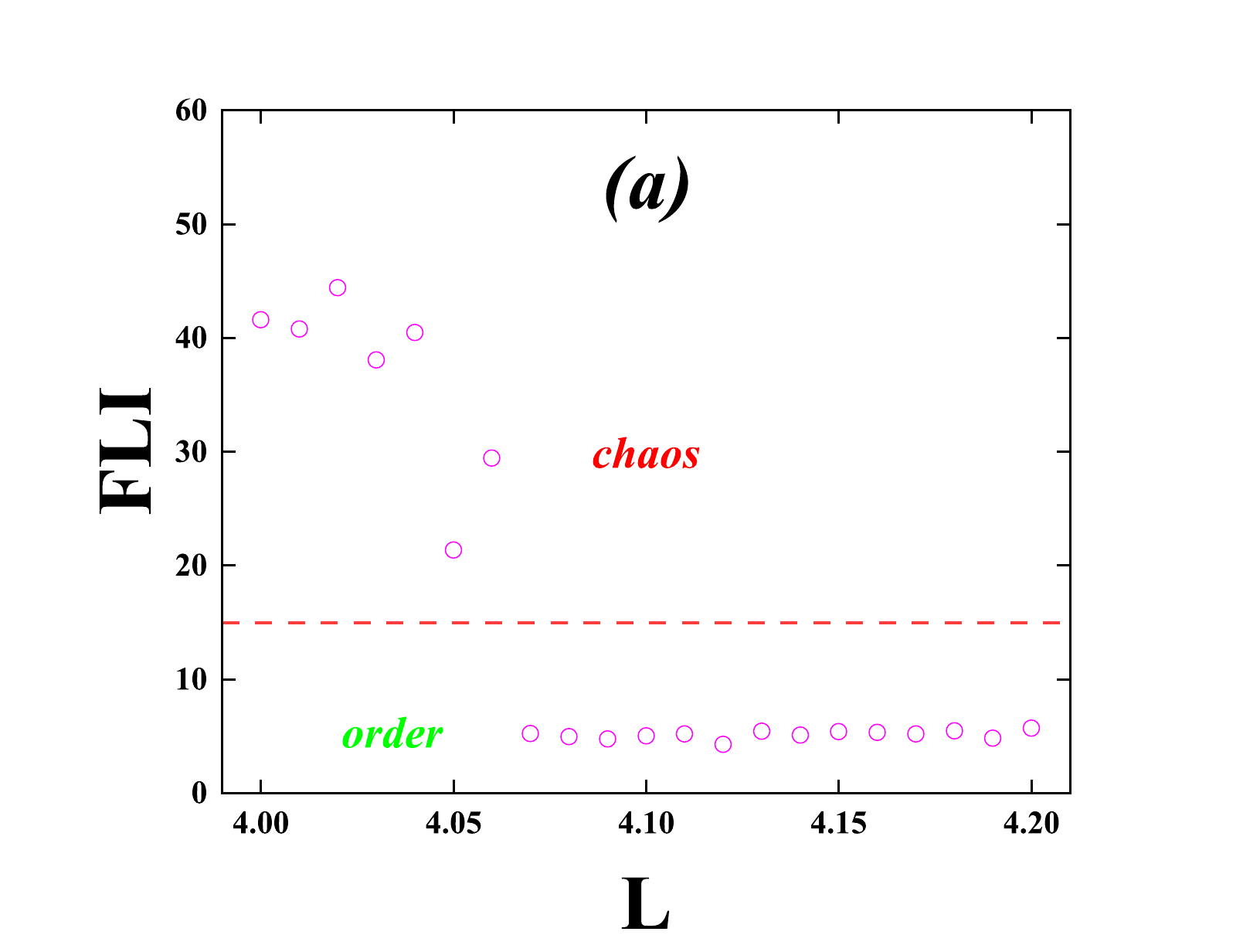}
\includegraphics[scale=0.3]{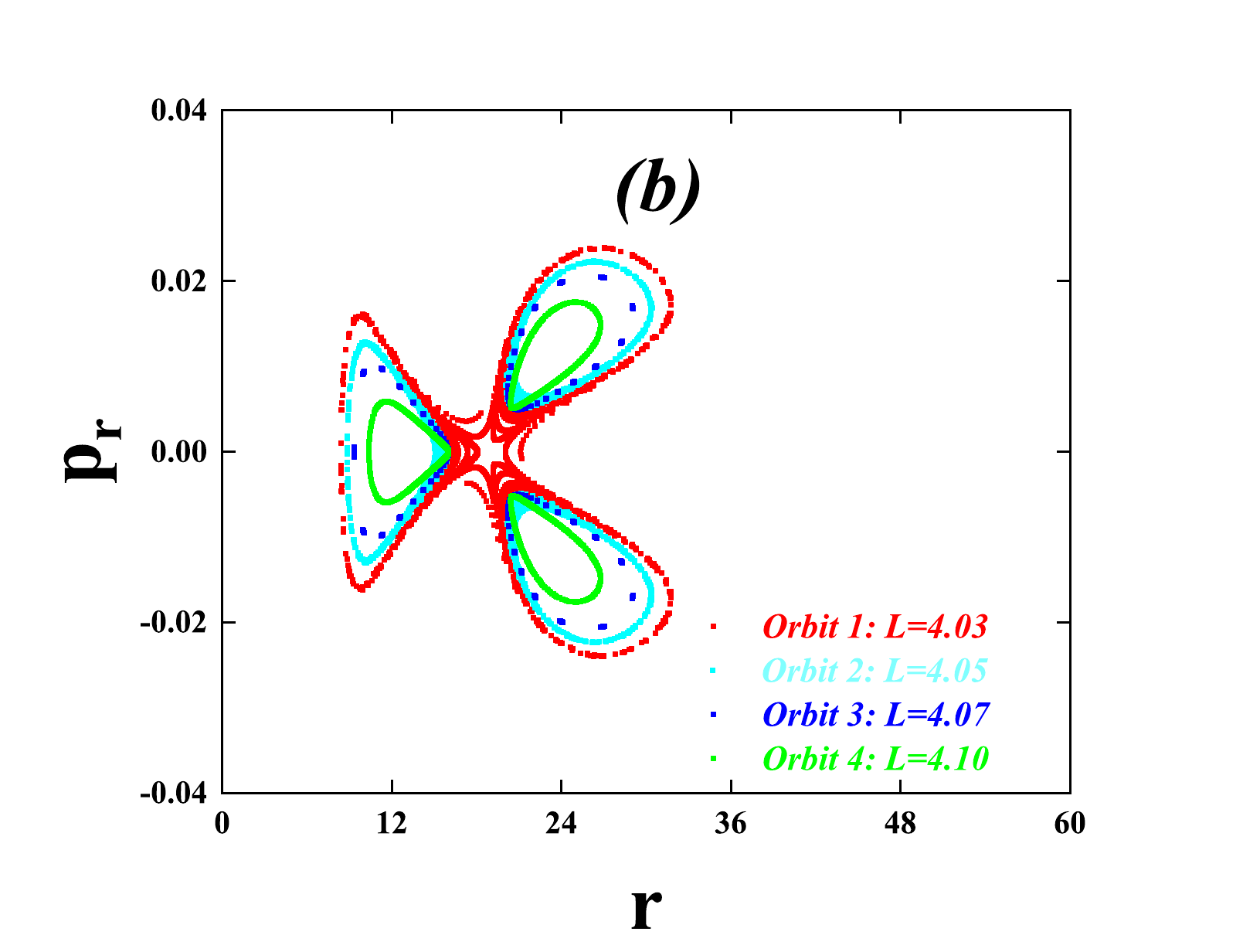}
\includegraphics[scale=0.5]{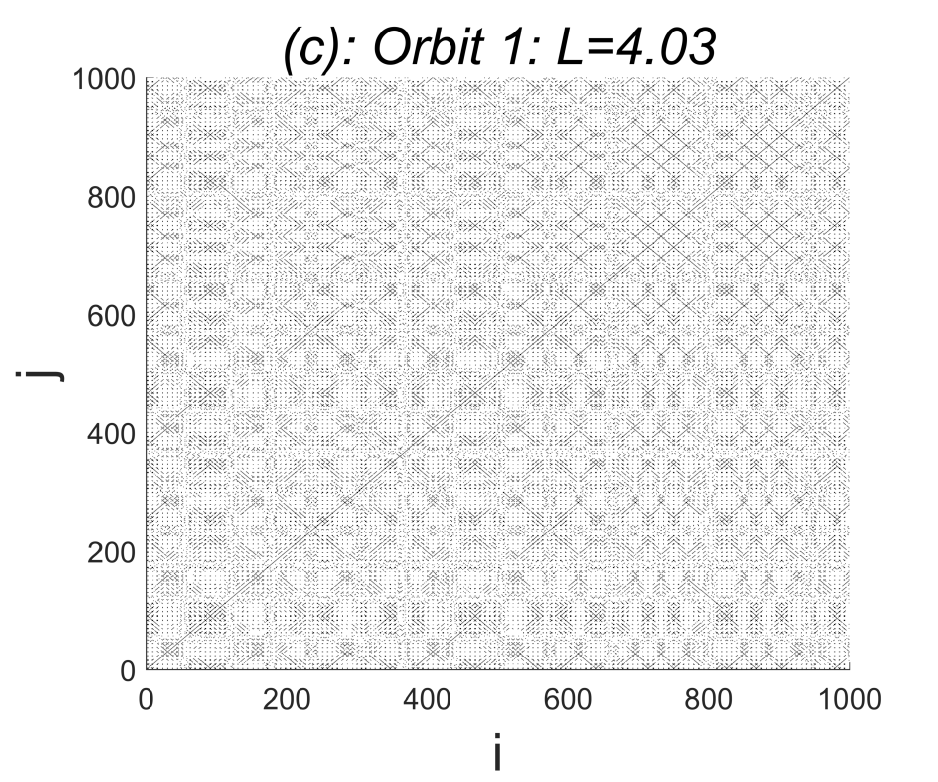}
\includegraphics[scale=0.5]{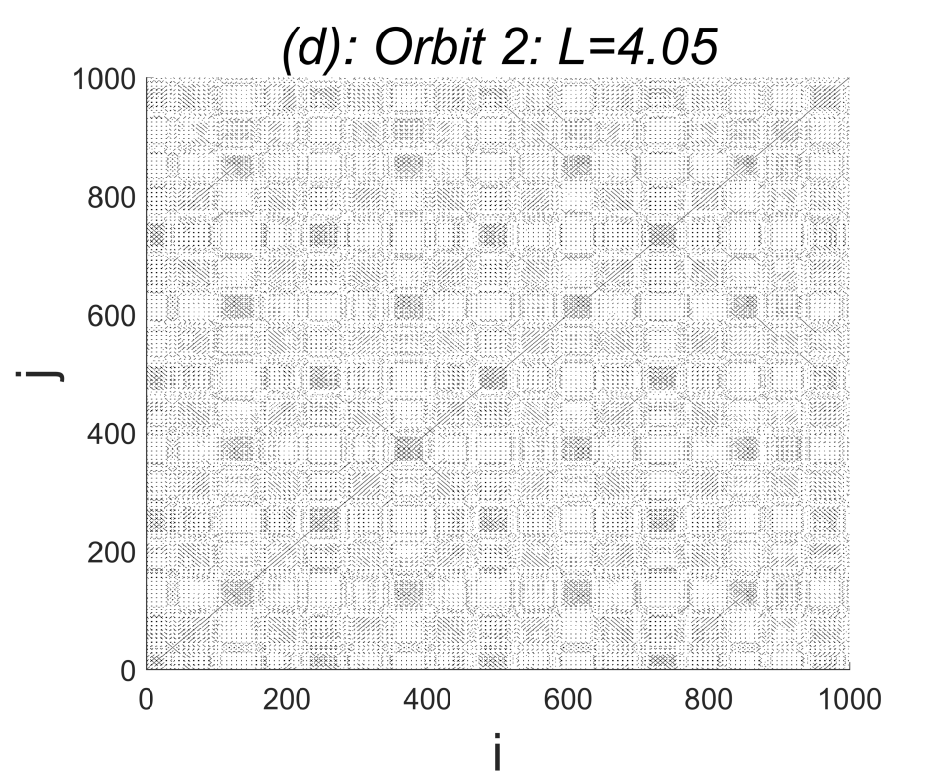}
\includegraphics[scale=0.5]{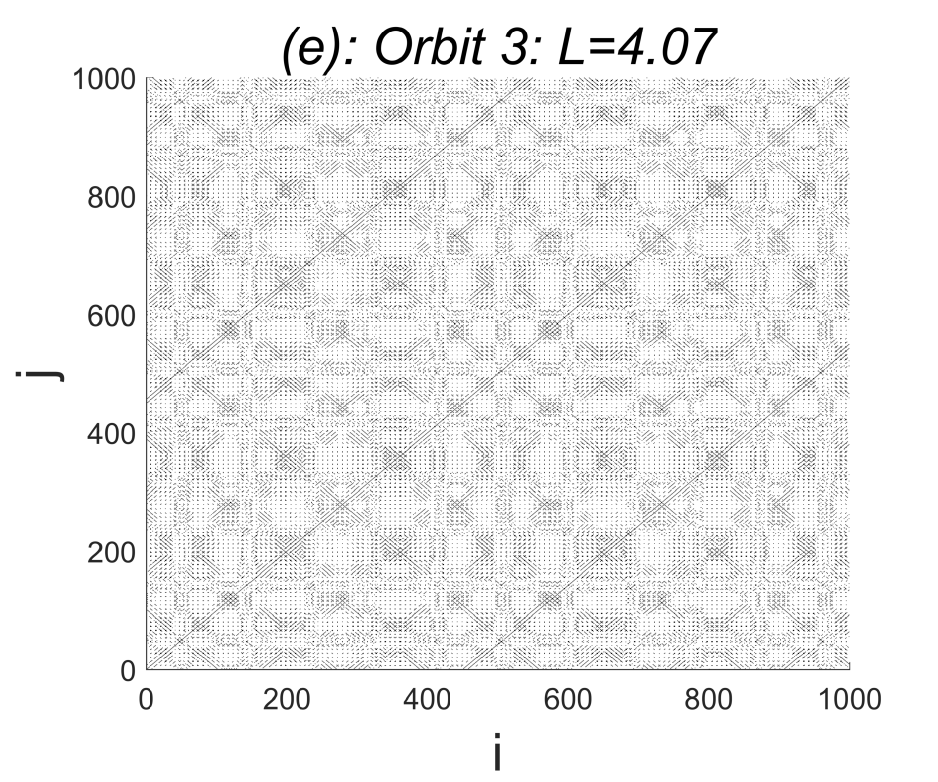}
\includegraphics[scale=0.5]{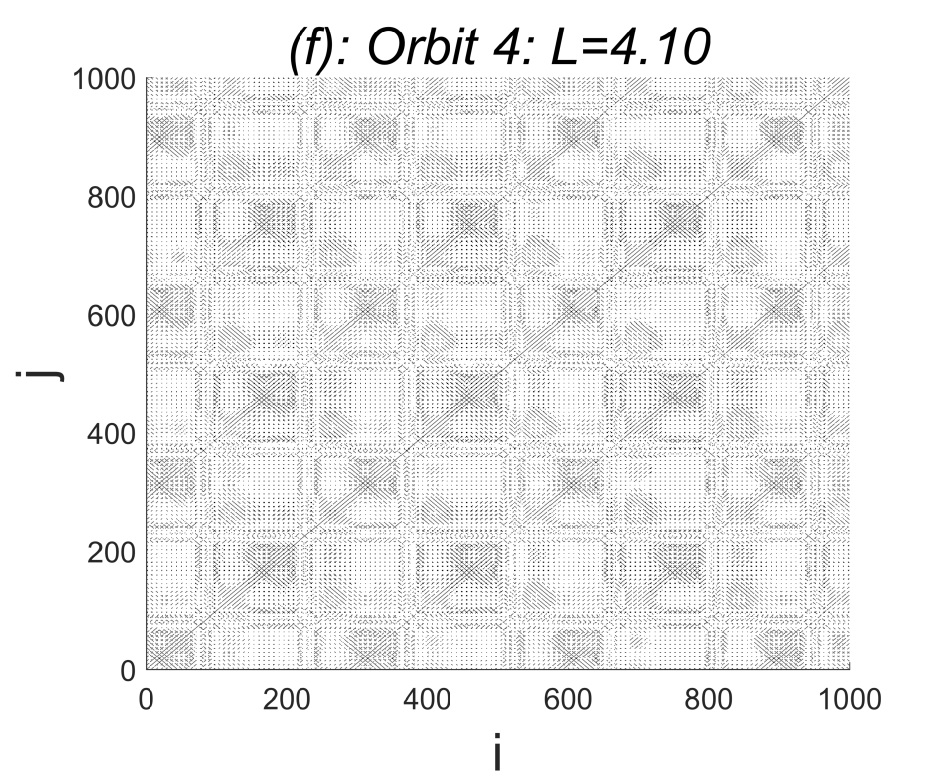}
\caption{(a) FLI describing a dynamical transition to chaos with
the angular momentum $L$ increasing, where the initial conditions
are $r=16$, $\theta=\pi/2$, and the other parameters are
$E=0.996$, $b=0.0006$, $\rho_\text{{eff}}=0.0001$. The transition to the regular dynamics from the chaotic dynamics
at $L=4.07$.  (b) Poincar\'{e} sections for four values of the angular momentum $L$.
(c)-(f): RPs for four values of the angular momentum $L$. The RPs
in (c) and (d) correspond to chaos, while the RPs in (e) and (f)
show the regular dynamics. }
 \label{Fig5}}
\end{figure*}

\begin{figure*}[htbp]
\center{
\includegraphics[scale=0.3]{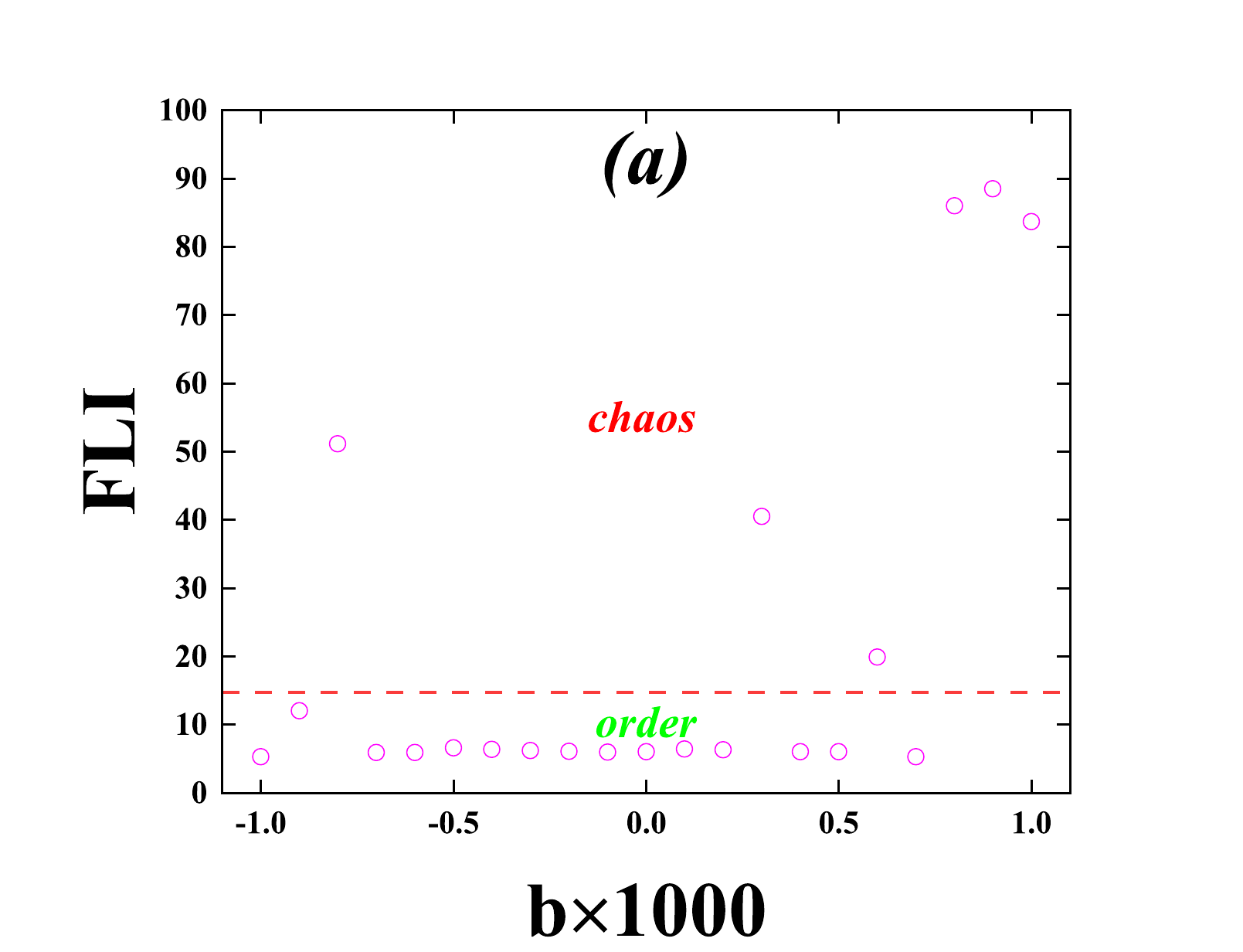}
\includegraphics[scale=0.3]{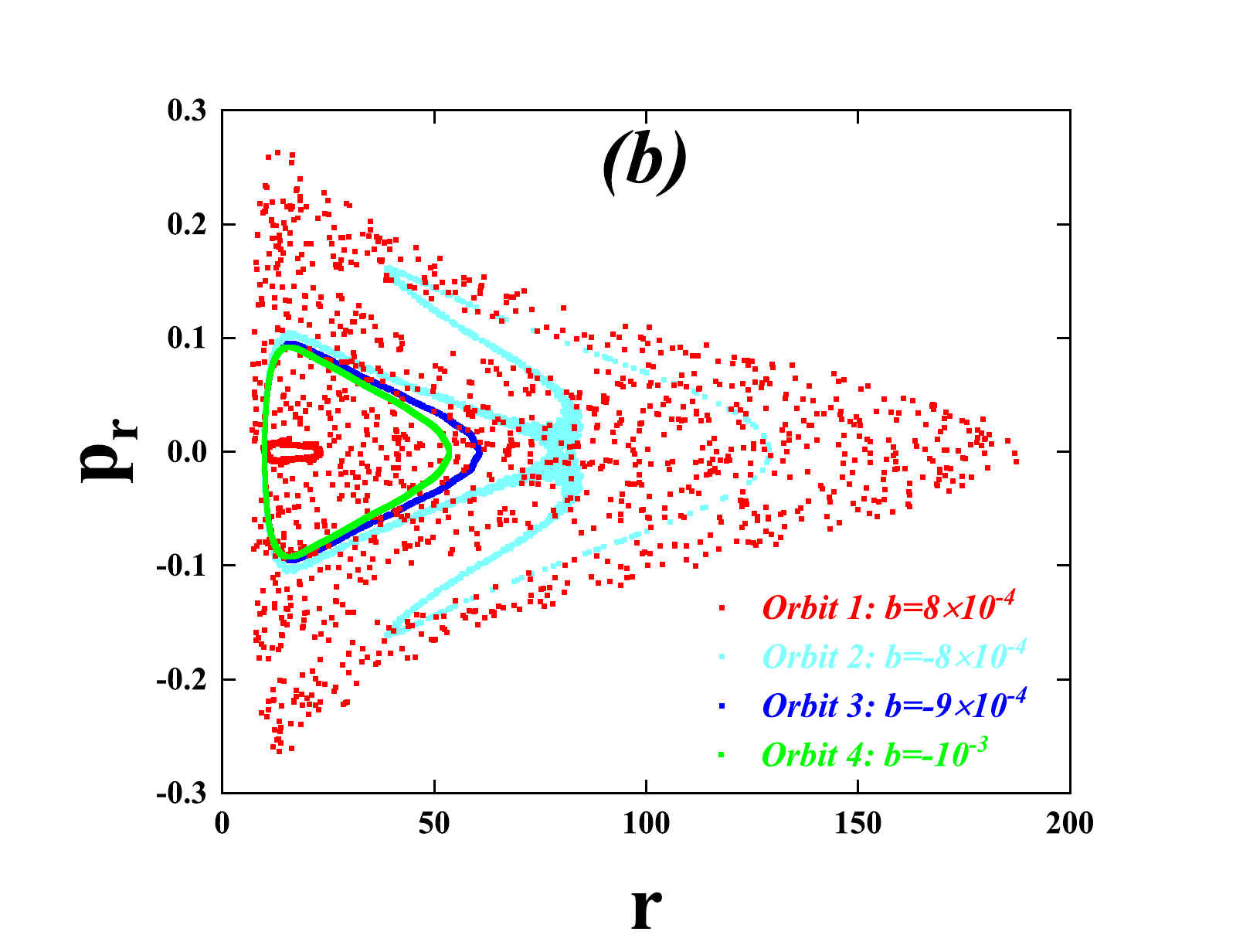}
\includegraphics[scale=0.5]{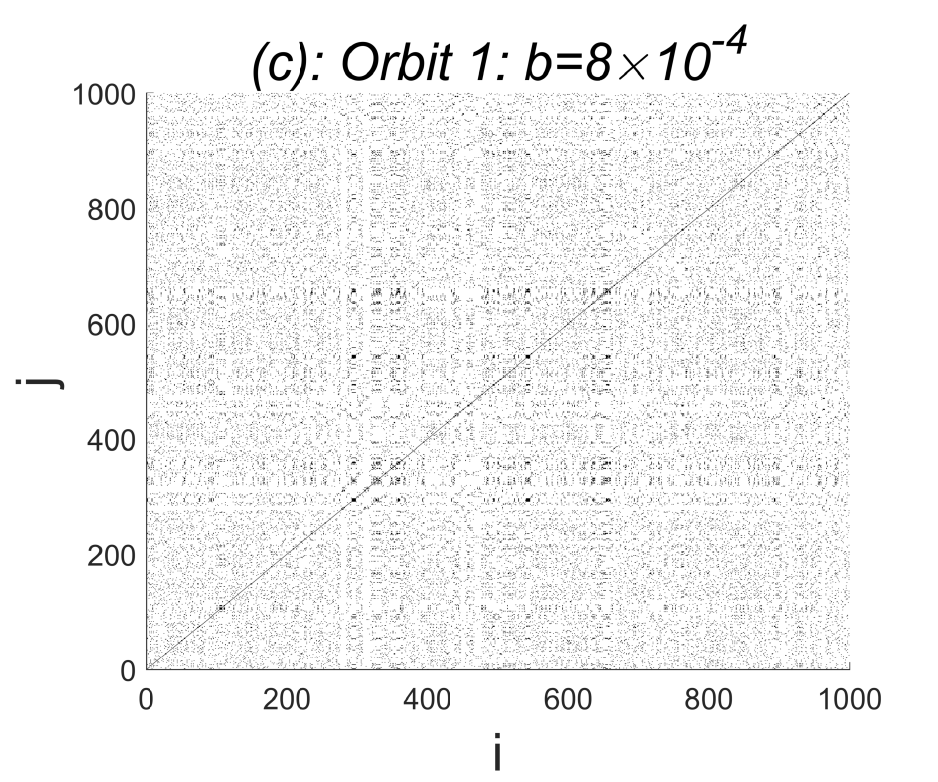}
\includegraphics[scale=0.5]{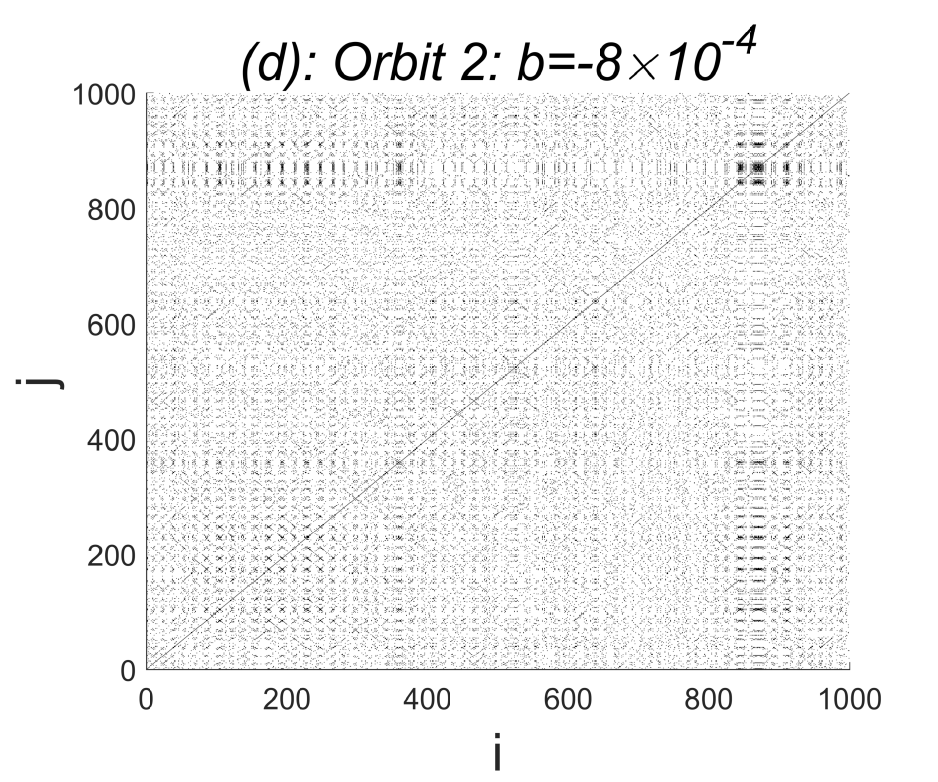}
\caption{(a) FLI describing a dynamical transition to chaos with
the magnetic parameter $b$ increasing, where the initial
conditions are $r=10$, $\theta=\pi/2$, and the other parameters
are $E=0.996$, $L=4.4$, $\rho_\text{{eff}}=0.0001$.
Chaos exists at $b=-0.0008,0.0003,0.0006$ and $0.0008\leq b\leq 0.001$.
(b) Poincar\'{e} sections for four values of the magnetic parameter
$b$. (c)-(f): RPs for four values of the magnetic parameter $b$.
The RPs in (c) and (d) correspond to chaos, while the RPs in (e)
and (f) show the regular dynamics.}
 \label{Fig6}}
\end{figure*}

\begin{figure*}[htbp]
\center{
\includegraphics[scale=0.3]{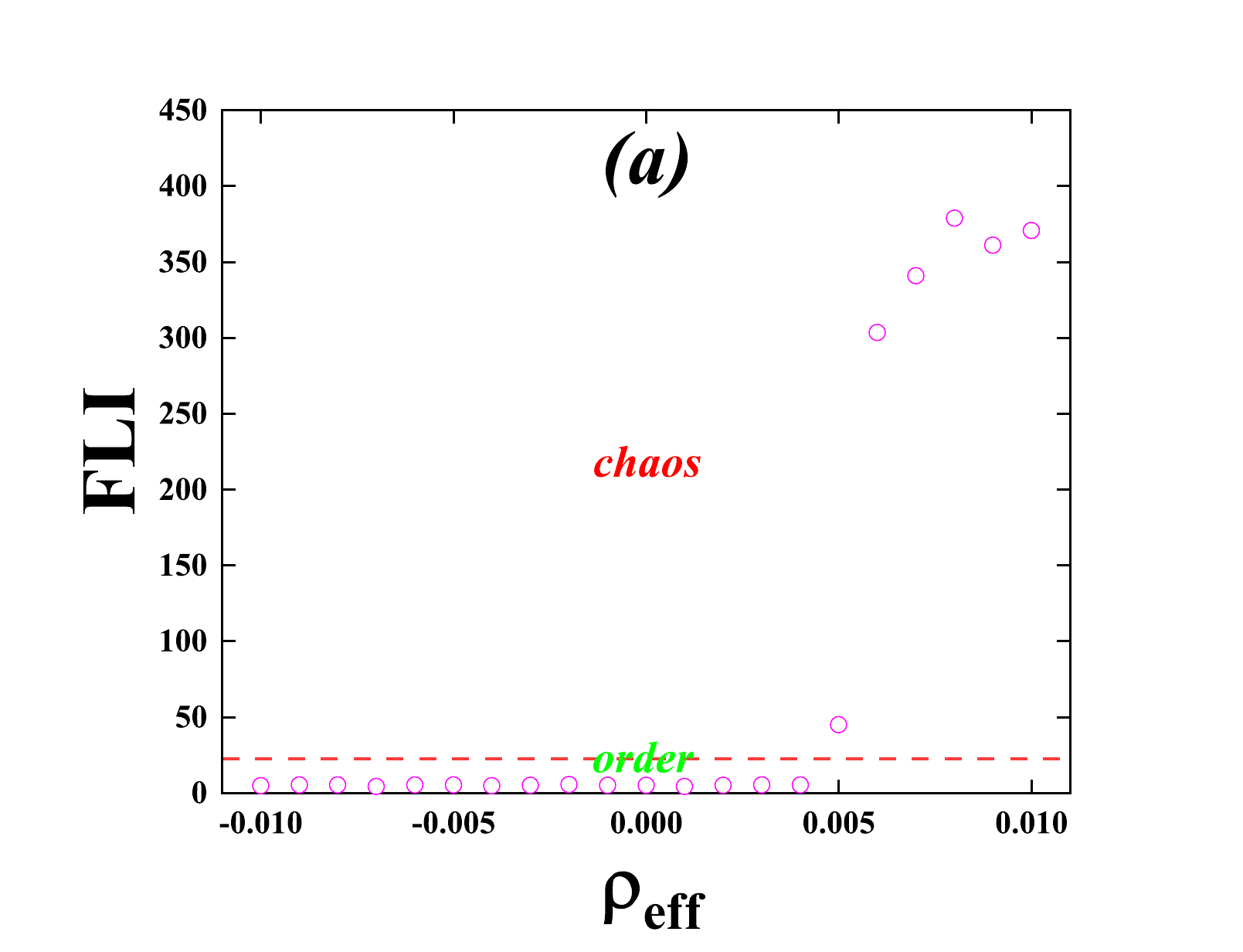}
\includegraphics[scale=0.3]{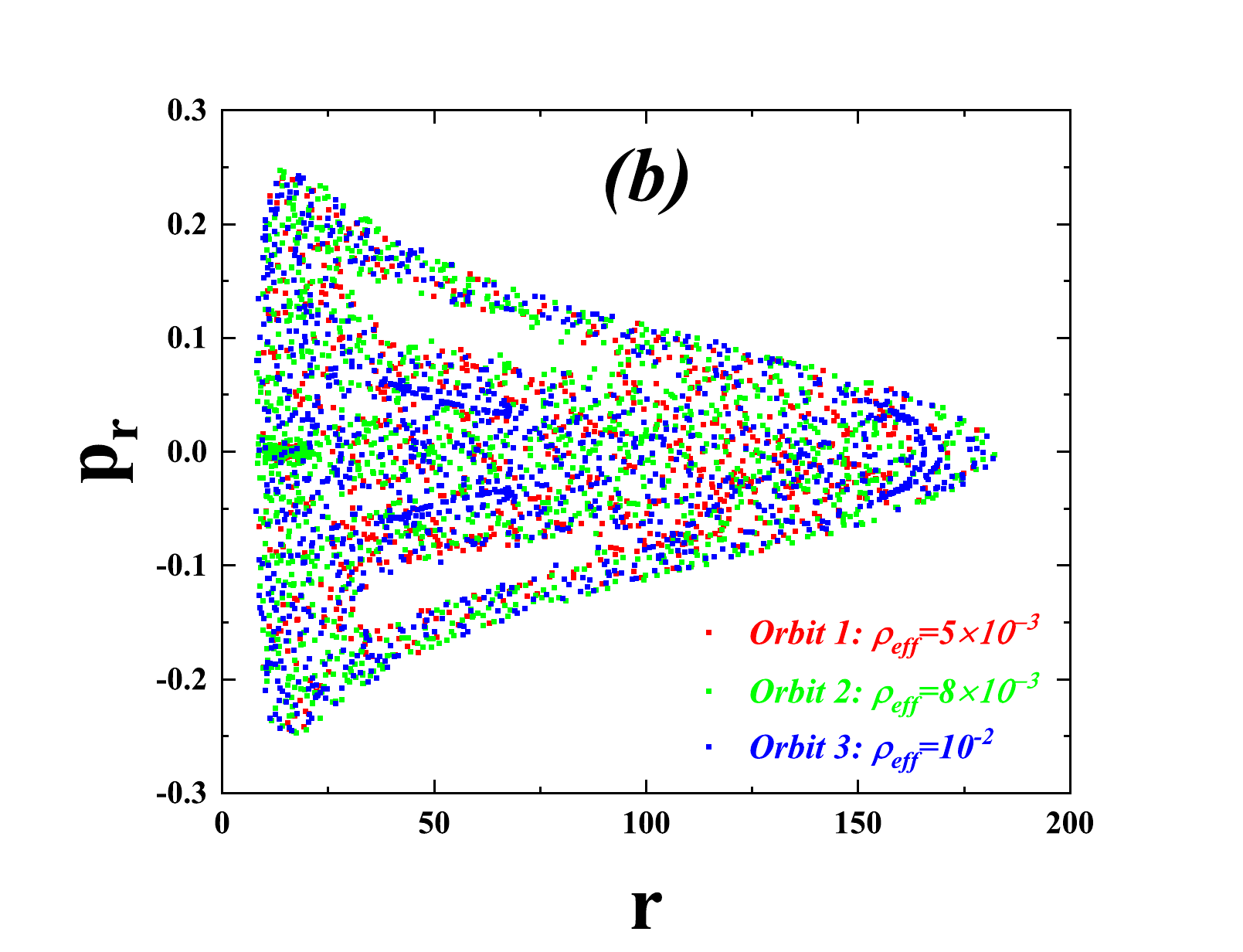}
\includegraphics[scale=0.5]{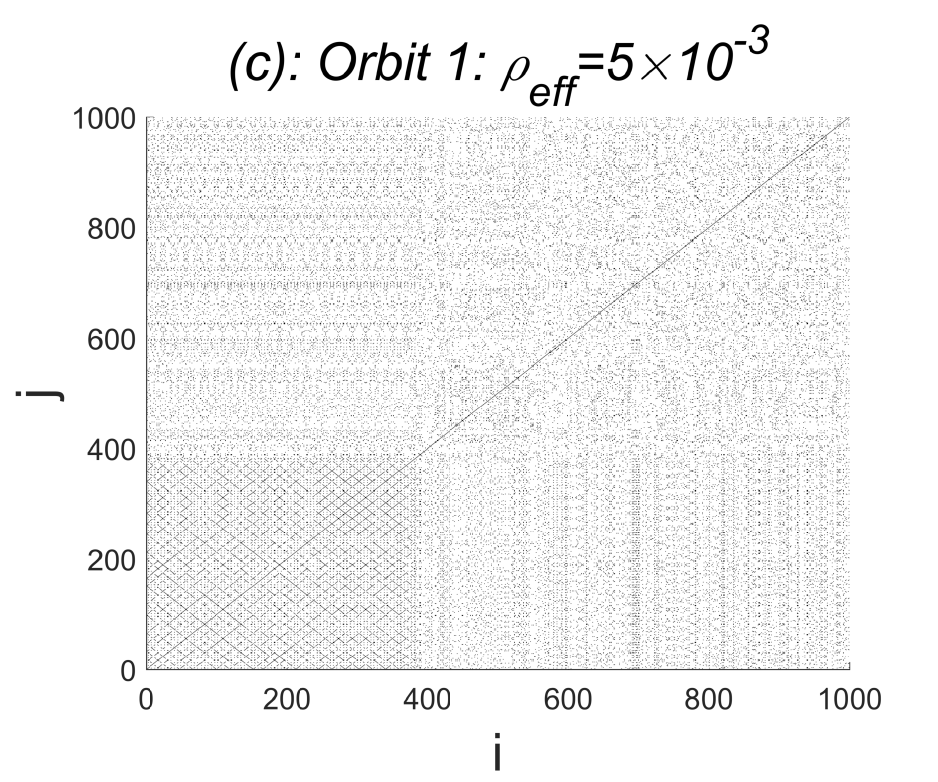}
\includegraphics[scale=0.5]{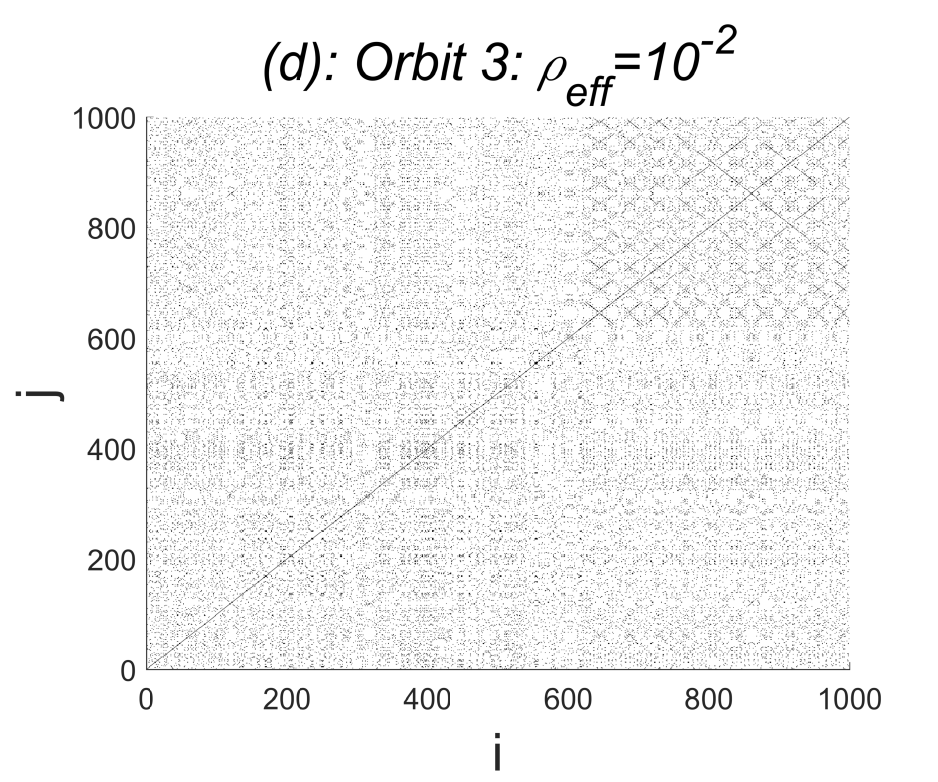}
\caption{(a) FLI describing a dynamical transition to chaos with
the hairy parameter $\rho_\text{{eff}}$ increasing, where the
initial conditions are $r=16$, $\theta=\pi/2$, and the other
parameters are $E=0.996$, $L=4.6$, $b=0.00088$. Chaos occurs when $\rho_\text{{eff}}\geq0.005$. (b) Poincar\'{e}
sections for three values of the hairy parameter
$\rho_\text{{eff}}$.  The RPs in (c) and (d) correspond to chaos.
 }
 \label{Fig7}}
\end{figure*}

\begin{figure*}[htbp]
\center{
\includegraphics[scale=0.3]{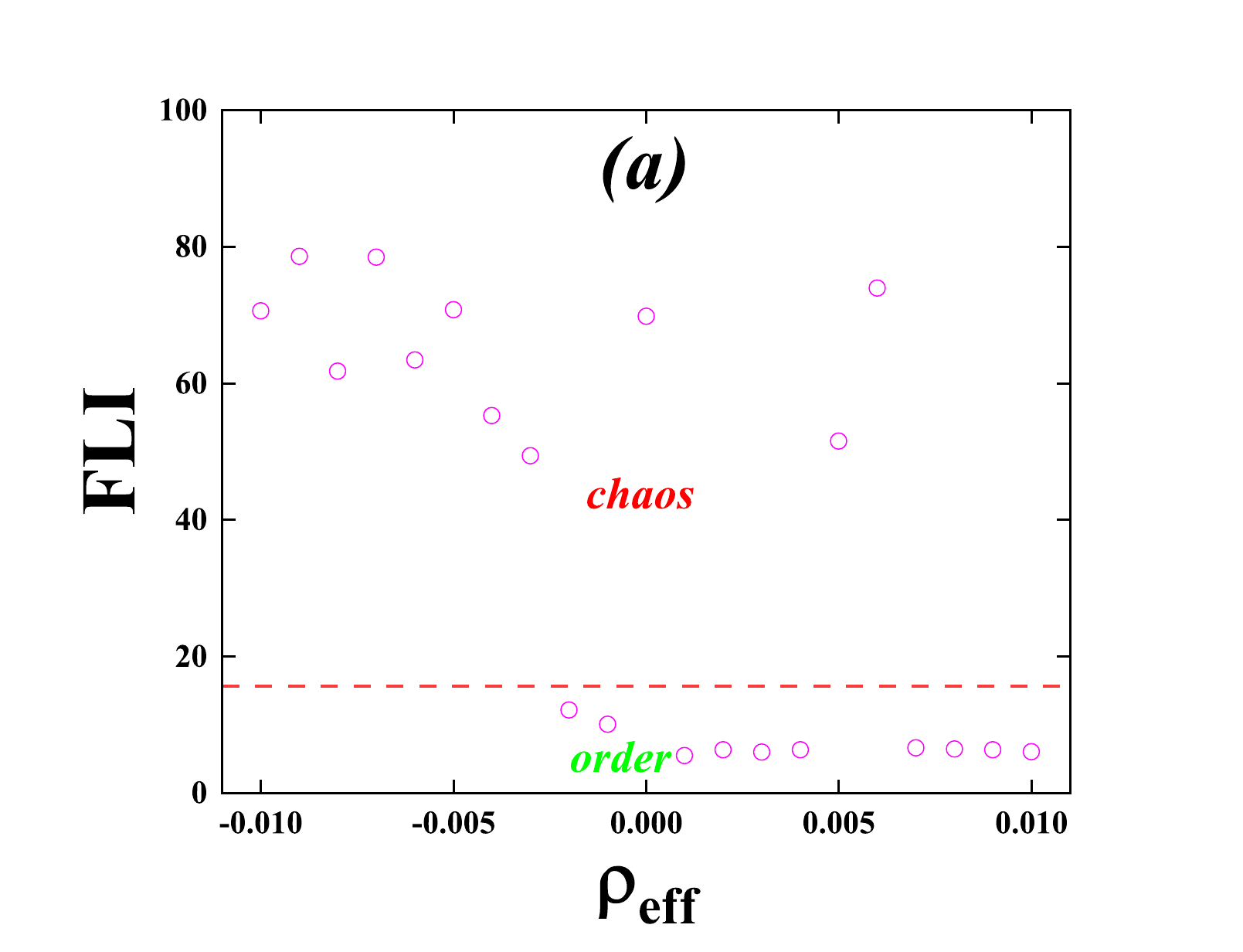}
\includegraphics[scale=0.3]{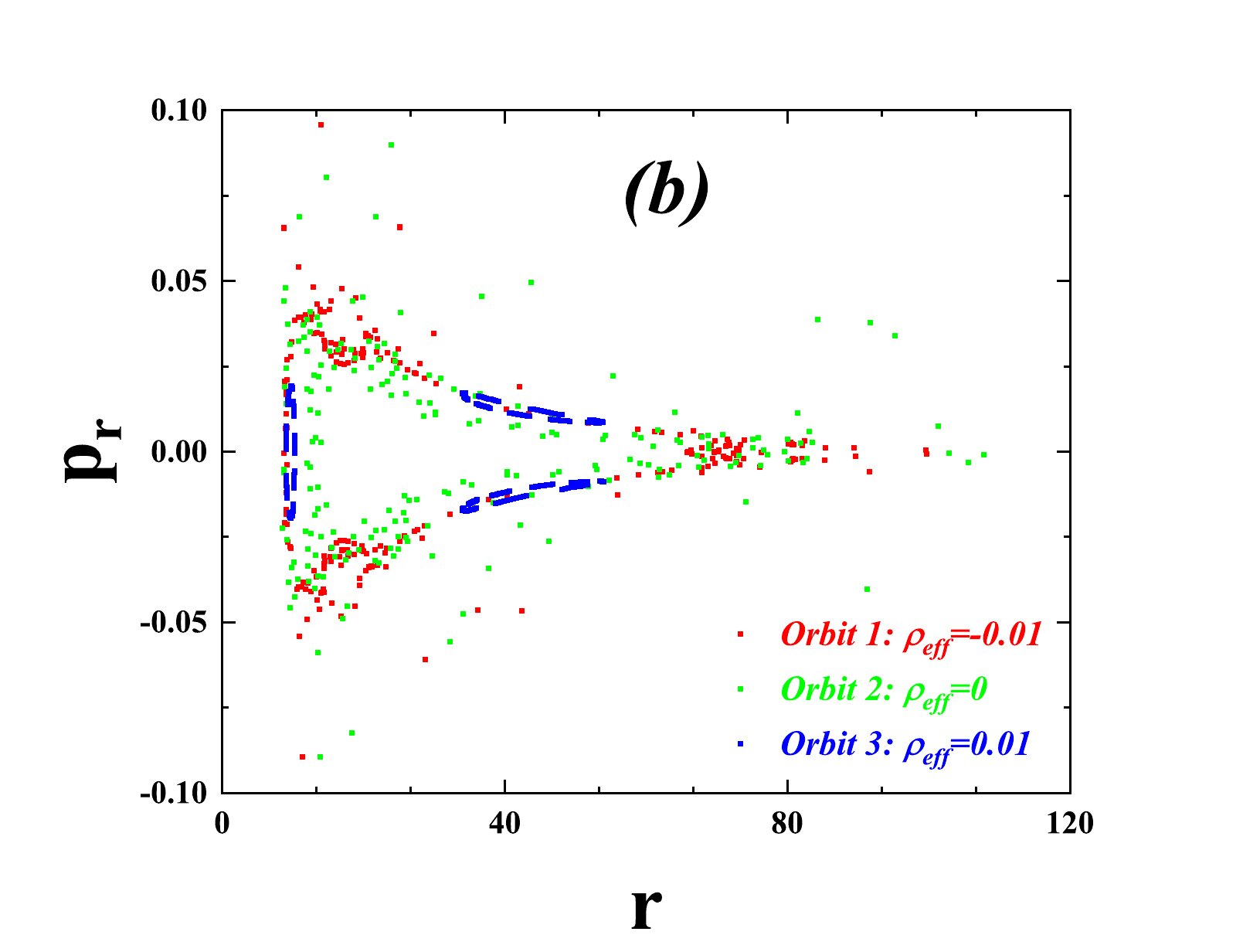}
\includegraphics[scale=0.5]{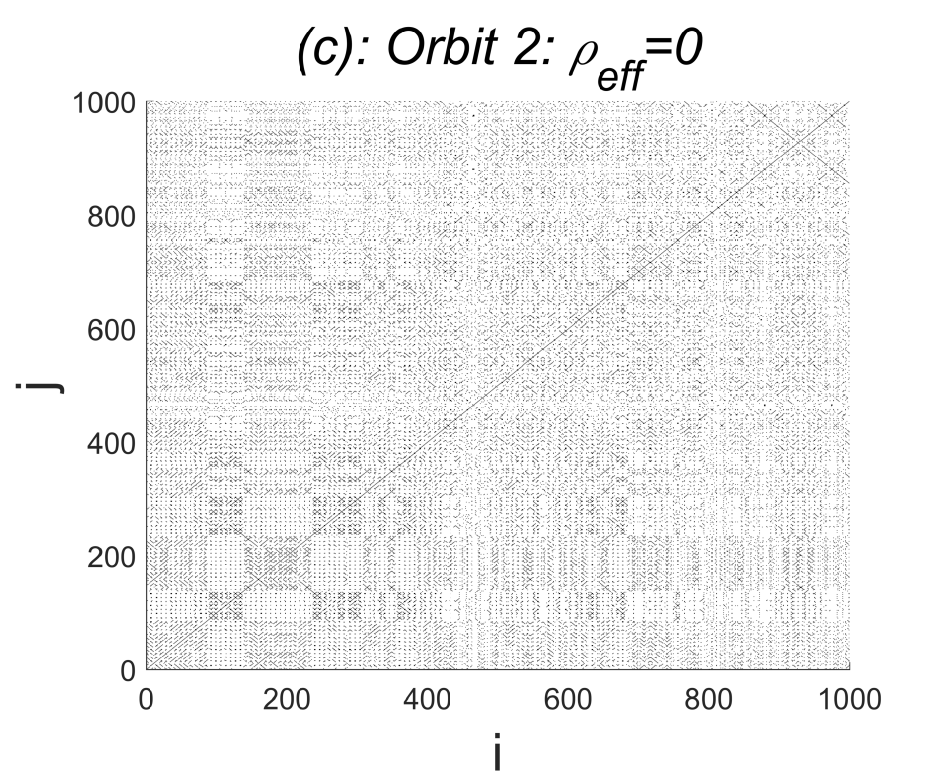}
\includegraphics[scale=0.5]{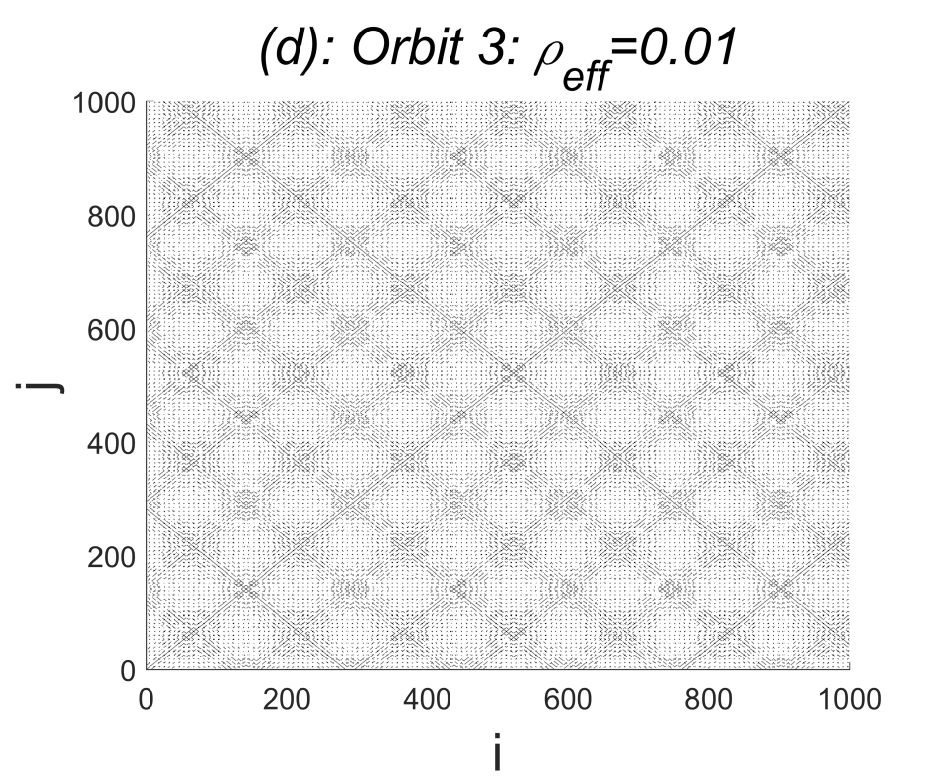}
\caption{Same as Fig. 7, but  the initial conditions are $r=9$,
$\theta=\pi/2$, and the other parameters are $E=0.999$, $L=4.6$,
$b=0.0001$. Chaos exists at $-0.01\leq \rho_{\text{eff}}
\leq-0.003$ and $\rho_{\text{eff}}=0,0.005,0.006$.
 }
 \label{Fig8}}
\end{figure*}

\begin{figure*}[htbp]
\center{
\includegraphics[scale=0.3]{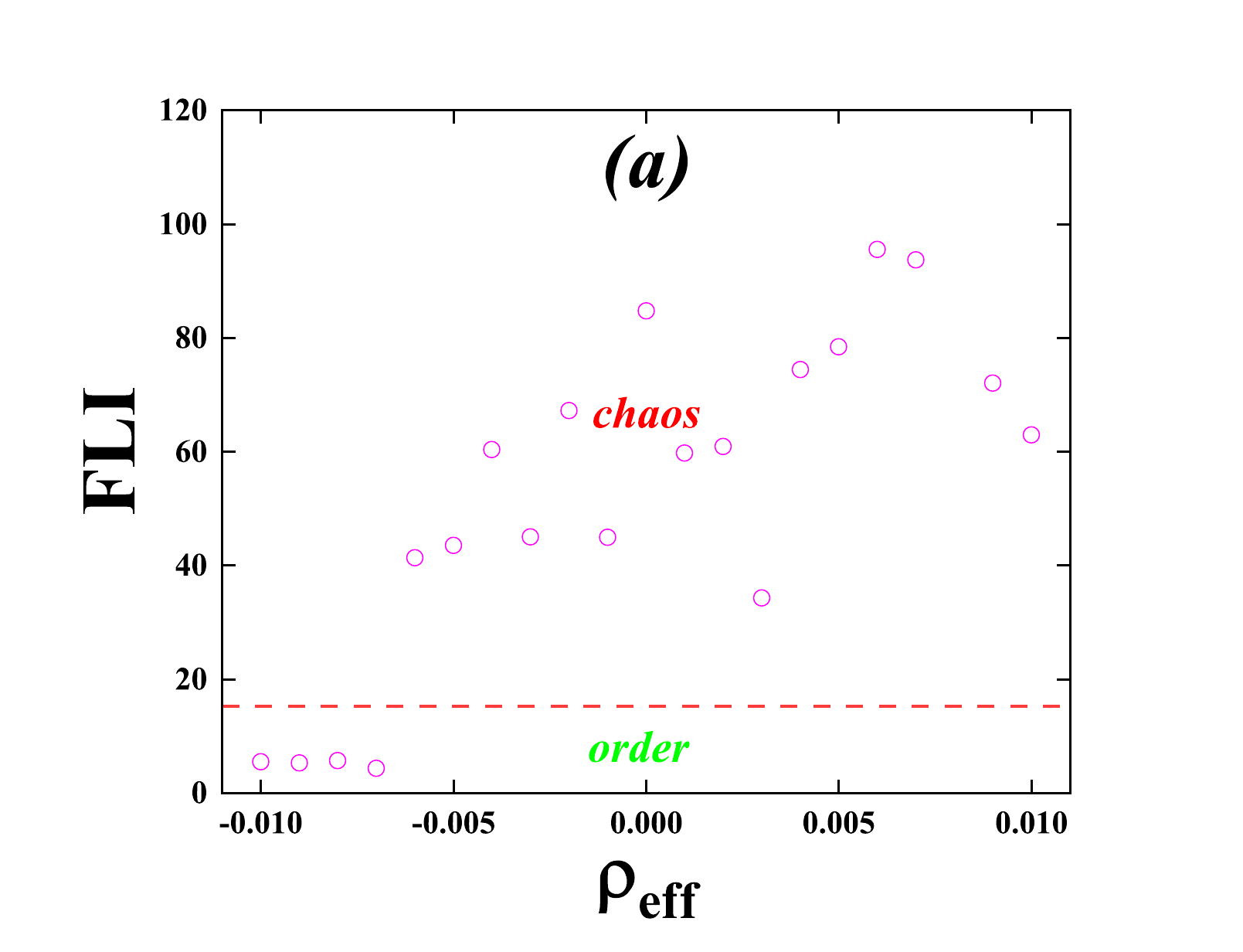}
\includegraphics[scale=0.3]{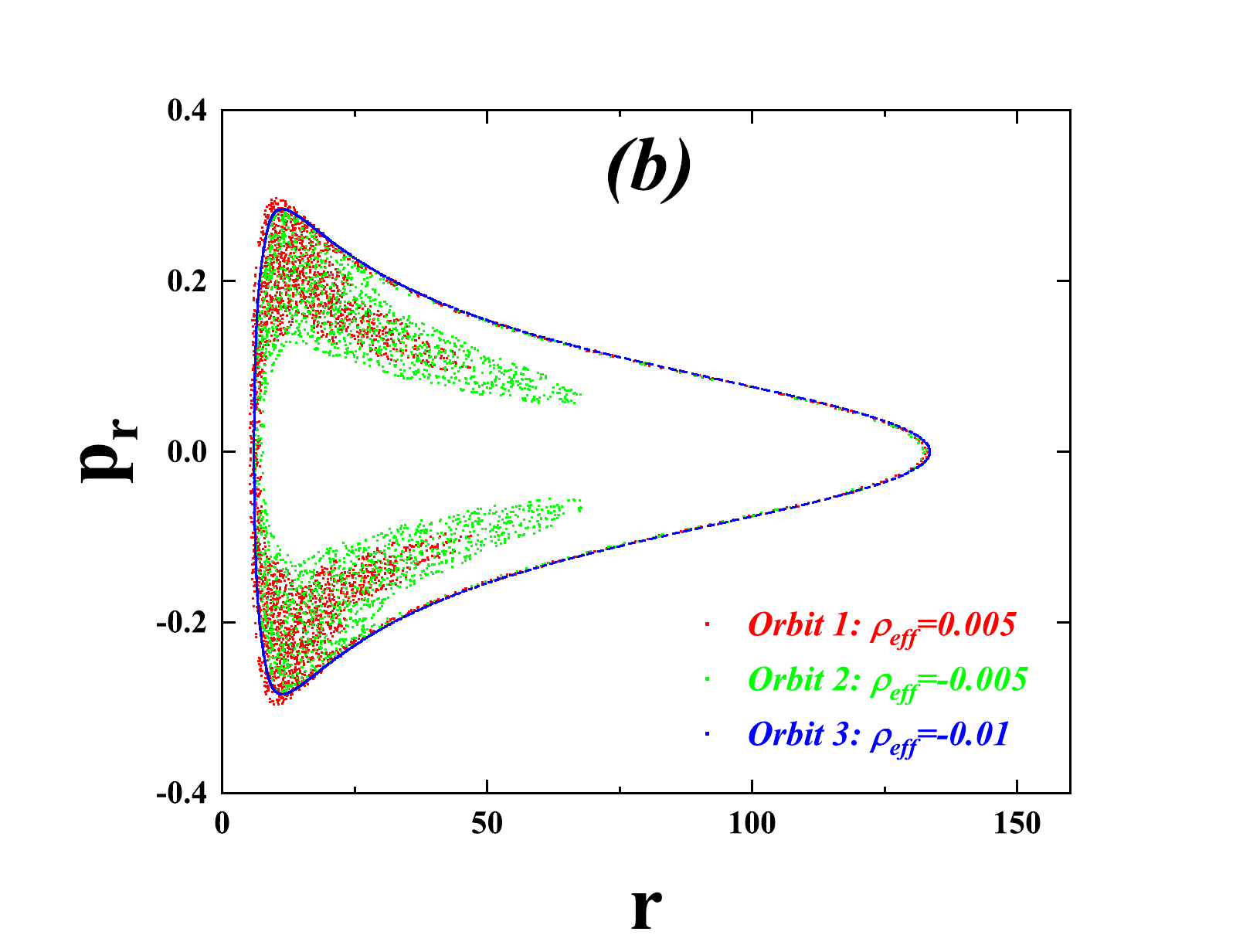}
\includegraphics[scale=0.5]{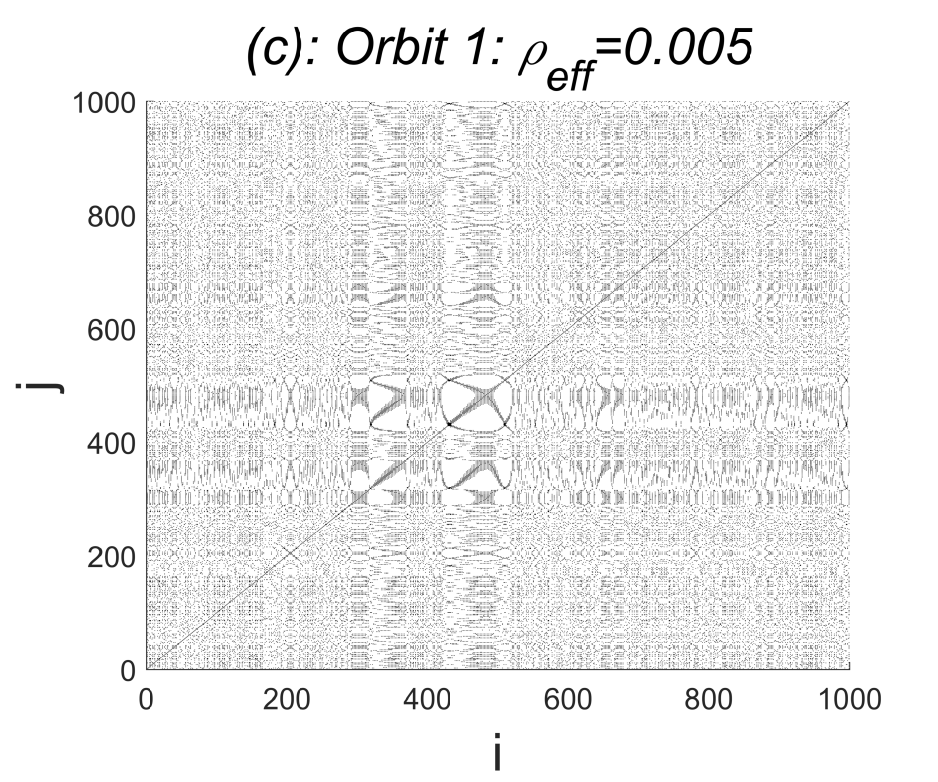}
\includegraphics[scale=0.5]{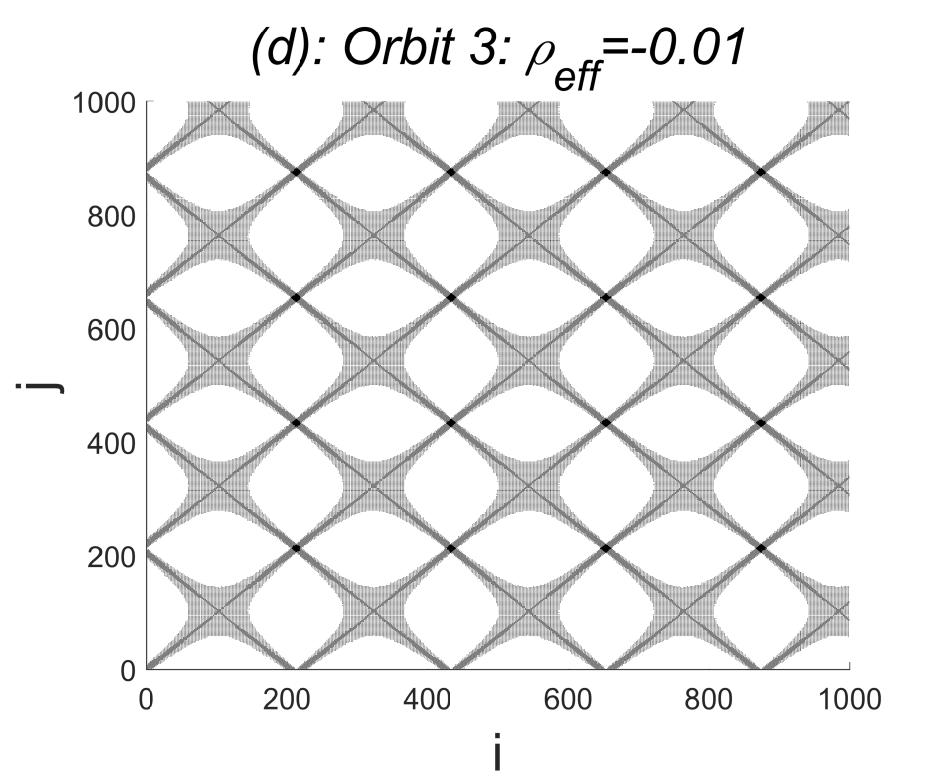}
\caption{Same as Fig. 7, but  the initial conditions are $r=6$,
$\theta=\pi/2$, and the other parameters are $E=0.996$, $L=4.1$,
$b=-0.0008$. There is the chaotic dynamics when $\rho_\text{{eff}}\geq-0.006$.  }
 \label{Fig9}}
\end{figure*}

\end{document}